\documentclass[12pt,preprint]{aastex}

\usepackage{color}
\usepackage[fleqn]{amsmath}
\usepackage{multirow}
\usepackage{multicol}
\usepackage{mathptmx}
\usepackage{graphicx}
\usepackage{bm}
\usepackage{mathrsfs}

\shorttitle{Effects of coupling constants} \shortauthors{Liu $\&$
Wu}

\begin{document}


\title{\large Effects of coupling constants on chaos of charged particles in the Einstein-\AE ther theory }

\author{Caiyu Liu$^{1,2}$, Xin Wu$^{1,2 \dag}$}
\affil{$^{1}$School of Mathematics, Physics and Statistics,
Shanghai University of Engineering Science, Shanghai 201620, China \\
$^{2}$ Center of Application and Research of Computational
Physics, Shanghai University of Engineering Science, Shanghai
201620, China} \email{$\dag$ Corresponding Author's
Email:wuxin$\_$1134@sina.com} 

\begin{abstract}

There are two  free coupling parameters $c_{13}$ and $c_{14}$ in
the Einstein-\AE ther metric describing a non-rotating black hole.
This  metric is the Reissner-Nordstr\"{o}m black hole solution
when $0\leq 2c_{13}<c_{14}<2$, but it is not for  $0\leq
c_{14}<2c_{13}<2$. When the black hole is immersed in an external
asymptotically uniform magnetic field, the  Hamiltonian system
describing the motion of charged particles around the black hole
is not integrable. However, the Hamiltonian allows for the
construction of explicit symplectic integrators. The proposed
fourth-order explicit symplectic scheme is used to investigate the
dynamics of charged particles because it exhibits excellent
long-term performance in conserving the Hamiltonian. No universal
rule can be given to the dependence of regular and chaotic
dynamics on varying one or two parameters $c_{13}$ and $c_{14}$ in
the two cases of $0\leq 2c_{13}<c_{14}<2$ and  $0\leq
c_{14}<2c_{13}<2$. The distributions of order and chaos in the
binary  parameter space $(c_{13},c_{14})$ rely on different
combinations of the other parameters and the initial conditions.

\end{abstract}




\section{Introduction}

Detections of gravitational waves [1] and event-horizon-scale
images of M87* [2] have strongly supported the prediction of
Einstein's general theory of relativity on the existence of black
holes. Although the standard general relativity has attained the
great success, its extension and development are still necessary
because of its limits. One example is that the dark universe
beyond the Einstein's theory may provide a good explanation for
the apparent accelerating expansion of the Universe. Another
example is the difficulty of building a complete theory unifying
interactions and particles in the general relativity and quantum
mechanics. Such theories of gravity are called as extended,
alternative or modified gravitational theories.

There are numerous modified gravitational theories in the
literature. Some of them are scalar-tensor theories of gravity
[3,4], scalar-tensor-vector theories [5,6], Ho\v{r}ava-Lifschitz
gravity [7], Kaluza-Klein theories of gravity [8,9], $F(R)$
gravity [10], quantum field theory in curved spacetime [7,11], and
Einstein-\AE ther theories [12-14]. As far as the Einstein-\AE
ther theories are concerned, they are obtained by coupling the \AE
ther field with other fields, such as an electromagnetic field.
They violate the Lorentz invariance that is one of the fundamental
principles of Einstein's general relativity. The so-called \AE
ther means  the presence of Lorentz-violating vector fields. This
particular property can dramatically affect cosmology, and can
even affect the growth rate of structure in the Universe. See
[15-17] for more information on these modified gravitational
theories.

From the astrophysical perspective, it is interesting to study the
motion of test particles around a black hole in the standard
general relativity or a certain modified theory of gravity. The
radiation emission by the particles with high energies in the
accretion disk leads to the electromagnetic spectrum of
astrophysical black hole. The spectrum is useful to understand the
effects of Doppler and gravitational red-shift. Circular orbits
and the innermost stable circular orbits (ISCOs) of the particles
near the black hole can give some important astrophysical
information on the nearby activity of the black hole. Thus, their
studies have appeared in a large number of publications (see,
e.g., [18-28]). The study of photon circular orbits also has a
crucial application in astrophysics because  the existence of
photon circular orbits is closely related to the computation of
black hole shadows [29-31].  When an external asymptotically
uniform magnetic field is included in the vicinity of the black
hole, the dynamics of electrically charged particles is
nonintegrable in general and is even chaotic. In fact, there are
lots of works about the presence of chaotic motions  of charged
particles around black holes (see, e.g., [32-39]). The chaotic
charged particle dynamics is helpful for charged particle
acceleration along the magnetic field lines [40]. Chaotic photon
motions in non-integrable black hole spacetimes also give
self-similar fractal structures to the black hole shadows [41,42].

Rayimbaev et al. [14] studied the effects of Einstein-\AE ther
gravity on the dynamics of magnetized particles around a black
hole surrounded by an external asymptotically uniform magnetic
field in the equatorial plane. They mainly discussed the impacts
of varying the free coupling parameters $c_{13}$ and $c_{14}$ of
the Einstein-\AE ther  theory on the radii of the ISCOs,  the
location of circular orbits, the amount of center-of-mass energy
and the strength of the magnetic field. Unlike these authors, we
consider in the present work how the two coupling parameters
affect the dynamics of electrically charged particles in the
3-dimensional space rather than the equatorial plane. In addition,
we design explicit symplectic integrators for the Hamiltonian
system describing the motion of charged particles near the
Einstein-\AE ther black hole by following the recent works
[43-47]. Such symplectic methods that maintain symplectic nature
of Hamiltonian dynamics are the most appropriate solvers for
studying the long-term evolution of Hamiltonian systems [48]. One
of the proposed explicit symplectic integrators combined with
chaos indications such as fast Lyapunov indicators (FLIS) [49] is
used to explore the dependence of regular and chaotic dynamics on
varying one or two parameters $c_{13}$ and $c_{14}$.

The paper is organized as follows. In Section 2, we introduce the
Einstein-\AE ther black hole metric and its corresponding
Hamiltonian system. In Section 3, we establish explicit symplectic
integrators and investigate the regular and chaotic dynamics of
charged particles. Finally, results are summarized in Section 4.

\section{Einstein-\AE ther black hole metric}

Based on the action of Einstein-\AE ther theory,
    a metric describing a non-rotating black hole [14] is written in the
 Schwarzschild coordinates  $(t, r,\theta,\varphi)$ as
    \begin{eqnarray}
        ds^2 &=& g_{\alpha \beta }dx^\alpha dx^\beta=-fdt^2+f^{-1}dr^2+r^{2}(d\theta ^2+\sin^2\theta d\varphi ^2), \\
        f(r) &=& 1-\frac{2M}{r}\left(1+\frac{2c_{13}-c_{14}}{4(1-c_{13})}\frac{M}{r}\right).
    \end{eqnarray}
Here, $t$ is a coordinate time, and $M$ is the mass of the black
hole. $c_{13}=c_{1}+c_{3}$ and $c_{14}=c_{1}+c_{4}$ are two
coupling constants of the Einstein-\AE ther theory. Of course,
    $c_1$, $c_3$, $c_4$ and the following parameter $c_2$ are still coupling constants. In fact, they
    are some of the coefficients in the tensor
    \begin{eqnarray}
        M^{\alpha\beta}_{\mu\nu}=c_1g_{\mu\nu}g^{\alpha\beta}+c_2\delta^{\alpha}_{\mu}\delta^{\beta}_{\nu}
        +c_3\delta^{\alpha}_{\nu}\delta^{\beta}_{\mu}-c_4u^{\alpha}u^{\beta}g_{\mu\nu},
    \end{eqnarray}
    which determines the Lagrangian of \AE ther field
    \begin{eqnarray}
        \mathcal{L}_{ \ae}=-M^{\alpha\beta}_{\mu\nu}(D_{\alpha}u^{\mu})
        (D_{\beta}u^{\nu})+\Lambda(g_{\mu\nu}u^{\mu}u^{\nu}+1)
    \end{eqnarray}
    in the action
    \begin{eqnarray}
        S_{\ae}=\frac{1}{16\pi G_{\ae}}\int d^4x
        \sqrt{-g}\left(\mathcal{R}+\mathcal{L}_{\ae}\right).
    \end{eqnarray}
    $G_{\ae}=G_N/(1-c_{14}/2)$ is the \AE ther gravitational constant,
    where $G_N$ stands for  the Newtonian gravitational constant.
    $D_{\alpha}$ denotes the covariant derivative with respect to
    $x^{\alpha}$, $\Lambda$ is the Lagrangian multiplier, and
    $u^{\alpha}$ is the \AE ther four-velocity. $g=|g_{\mu\nu}|$, and
    $\mathcal{R}$ is the curvature. When the coupling constants $c_{13}$ and $c_{14}$ are given different values, Equation (1)
represents different spacetimes. Here, three cases are listed.
Case (i): if $c_{14}=2c_{13}\neq 2$, Eq.
    (1) is the Schwarzschild metric.  Case (ii): when $0\leq c_{13}<1$ and
    $2c_{13}<c_{14}<2$, Eq. (1) represents the Reissner-Nordstr\"{o}m (RN)
    black hole with the electric charge
    \begin{eqnarray}
        Q=M\sqrt{(c_{14}-2c_{13})/[2(1-c_{13})]}
    \end{eqnarray}
    and two horizons
    \begin{eqnarray}
        R^{\pm}_h=M\left(1\pm\sqrt{(2-c_{14})/[2(1-c_{13})]}\right).
    \end{eqnarray}
Case (iii): if $0\leq c_{13}<1$ and $0\leq c_{14}<2c_{13}<2$,
Equation (1) shows that the black hole with one horizon $R^{+}$
has no electric charge and is not the RN black hole. In the three
cases, the coupling constants $c_{13}$ and $c_{14}$ satisfy the
constraints $0\leq c_{13}<1$ and $0\leq c_{14}<2$ from theoretical
and observational bounds [14], which correspond to the presence of
one or two horizons. The speed of light $C$ and Newton's constant
of gravity $G_N$ are given geometrized units, $C = G_N= 1$.

Consider that an external asymptotically uniform magnetic field
exists  in the vicinity of black hole. This magnetic field is
static, axially symmetric and homogeneous at the spatial infinity,
and has its strength $B>0$ at the spatial infinity. It is so weak
that it does not affect the spacetime geometry. Based on the
Wald's method, only one non-vanishing component of the
electromagnetic four potential was given in Ref. [14] by
\begin{eqnarray}
    A_\varphi =\frac{1}{2}Br^2\sin^2\theta.
\end{eqnarray}
Although such a weak magnetic field exerts a negligible influence
on the spacetime geometry, it plays an important role in the
motion of a charged test particle with mass $m$ and charge $q$.
The motion is described by the following Hamiltonian
\begin{eqnarray}
    H=\frac{1}{2m}g^{\mu \nu}(p_\mu -qA_\mu)(p_\nu-qA_\nu),
\end{eqnarray}
where the momentum $p_\mu$ satisfies the condition
\begin{eqnarray}
    \dot{x}^\mu=\frac{\partial H}{\partial p_\mu}=\frac{1}{m}g^{\mu \nu}(p_\nu-qA_\nu),
\end{eqnarray}
or an equivalent form
\begin{eqnarray}
p_\mu=mg_{\mu \nu}\dot{x}^\nu+qA_\mu.
\end{eqnarray}
The four-velocity
$\dot{x}^\mu=(\dot{t},\dot{r},\dot{\theta},\dot{\varphi})$ is a
derivative of coordinate $x^\mu$ with respect to the proper time
$\tau$.

To simplify the expressions, we adopt dimensionless operations
through scale transformations as follows: $t\to tM$, $\tau\to \tau
M$, $r\to rM$, $B\to B/M$, $E\to mE$, $p_t\to mp_t$, $p_r\to
mp_r$, $L\to MmL$, $p_{\theta}\to Mmp_{\theta}$, $q\to mq$, and
$H\to mH$. In this case, $m$ and $M$ as two mass factors are
eliminated in the above expressions.

Because  the Hamiltonian (9) does not explicitly depend on the
coordinates $t$ and $\varphi$, $p_t$ and $p_\varphi$ are constants
of motion, which correspond to the energy $E$ and angular momentum
$L$ of the particle:
\begin{eqnarray}
p_t &=&-f\dot{t}=-E,\\
p_\varphi &=&r^2\sin^2 \theta \dot{\varphi }+q A_\varphi=L.
\end{eqnarray}
The Hamiltonian (9) is rewritten as
\begin{eqnarray}
    H   &=& T+V,\\
    T &=& \frac{p_{r}^2}{2}-\frac{p_{r}^2}{r}+\frac{p_{\theta }^2}{2r^2} -\frac{p_{r}^2(2c_{13}-c_{14})}{4r^2(1-c_{13})} , \\
V &=&\frac{L^2}{2r^2\sin^2\theta} +\frac{b^2r^2}{8}+\frac{bL}{2\sin\theta } \nonumber\\
    &&+\frac{E^2r^2(1+c_{13})}{2c_{13}-c_{14}+4(1-c_{13})r-2(1-c_{13})r^2},
\end{eqnarray}
where $b=Bq$. Noticing that the four-velocity $\dot{x}^\mu$ always
satisfies the relation $g_{\mu \nu}\dot{x}^\mu\dot{x}^\nu=-1$ for
a timelike geodesic, we have a third motion constant
\begin{eqnarray}
    H=-\frac{1}{2}.
\end{eqnarray}

\section{Numerical simulations}

At first, we introduce how to design  explicit symplectic
integrators for the Hamiltonian (14). Then, an appropriate one of
the integrators is used to study the regular and chaotic dynamics
of charged particles near the black hole.

\subsection{Explicit symplectic integrations}

As the authors of [44] claimed, the Hamiltonian (14) can be split
into five parts
\begin{eqnarray}
    H=H_1+H_2+H_3+H_4+H_5
\end{eqnarray}
where these sub-Hamiltonian systems are
\begin{eqnarray}
    H_1&=& V,\\
    H_2&=&\frac{p_{r}^2}{2},\\
    H_3&=&-\frac{p_{r}^2}{r},\\
    H_4&=&\frac{p_{\theta }^2}{2r^2},\\
    H_5&=&\frac{p_{r}^2(c_{14}-2c_{13})}{4r^2(1-c_{13})}.
\end{eqnarray}
The five splitting parts are solved analytically and their
analytical solutions are explicit functions of the proper $\tau$.
Operators for analytically solving these sub-Hamiltonians are
$\mathcal{H}_1$, $\mathcal{H}_2$, $\mathcal{H}_3$, $\mathcal{H}_4$
and $\mathcal{H}_5$ in turn. The five-splitting-part method is
based on the exclusion of Case (i) with $c_{14}=2c_{13}$.

Setting $h$ as a time step, we have a second-order explicit
symplectic algorithm for the Hamiltonian (14)
\begin{eqnarray}
    S_2(h)&=& \mathcal{H}_5(\frac{h}{2})\times\mathcal{H}_4(\frac{h}{2})\times
    \mathcal{H}_3(\frac{h}{2})\nonumber\times\mathcal{H}_2(\frac{h}{2})\times\mathcal{H}_1(h) \nonumber \\
    && \times\mathcal{H}_2(\frac{h}{2})\times\mathcal{H}_3(\frac{h}{2}) \times\mathcal{H}_4(\frac{h}{2})\times\mathcal{H}_5(\frac{h}{2}).
\end{eqnarray}
In such a construction, the sub-Hamiltonian $H_5$ is integrated by
advancing the time $h/2$. When the obtained solutions are taken as
the initial conditions, we solve the sub-Hamiltonian $H_4$ by
advancing the time $h/2$. The operation is also applied to the
sub-Hamiltonians $H_3$ and $H_2$. Taking the numerical solutions
at the time as the initial conditions,  we work out the
sub-Hamiltonian $H_1$ by advancing the time $h$. Then, we continue
to solve the sub-Hamiltonian $H_2$, $H_3$, $H_4$ and $H_5$ in
sequence by  advancing the time $h/2$. As a result, the numerical
solutions are outputted in one integration step. According to the
idea of Yoshida [50], the second-order explicit symplectic
integrator can be risen to a fourth-order algorithm
\begin{eqnarray}
    S_4=S_2(\gamma h)\times S_2(\delta h)\times S_2(\gamma h),
\end{eqnarray}
where $\gamma =1/(2-\sqrt[3]{2})$ and $\delta =1-2\gamma$. This
construction is explained as follows. We use the algorithm $S_2$
to act on the Hamiltonian (18) by advancing the time $\gamma h$.
Taking the numerical solutions as the initial conditions, we
continue to solve the Hamiltonian (18) by using the algorithm
$S_2$ in a span of time  $\delta h$. Finally, we repeat to use the
algorithm $S_2$ to integrate the Hamiltonian (18) by advancing the
time $\gamma h$. In this way, $\gamma h+\delta h+\gamma h=h$; that
is, one step computation finishes.

In our numerical tests, the fixed proper time step $h=1$ is used.
The parameters are given by $E=0.995$, $L=4.7$ and $b=9.1\times
10^{-4}$. The coupling constants are taken as $c_{13}=0.01$ and
$c_{14}=0.03$.  The initial conditions are $p_{r}=0$ and $ \theta
=\pi/2$. Given the initial separation $r$, the initial value of
$p_{\theta }(>0)$ is determined in terms of Equations (14)-(17).
The initial separations are $r=15$ for Orbit 1 and $r=35$ for
Orbit 2. When the two orbits are integrated by the second-order
method $S_{2}$ and the fourth-order method $S_{4}$, Hamiltonian
errors $\Delta H=H+1/2$ are plotted in Figure 1 (a) and (b). These
errors show no secular drift with time. This property is an
advantage of symplectic integrators. Of course, the errors are
sensitive dependence on  the step choice. They decrease with a
decrease of the time step. However, extremely small steps are not
permitted due to the increase of computational cost and the rapid
accumulation of roundoff errors. Relatively large steps are not,
either, because they lead to large errors and make the algorithms
unstable. The time step $h=1$ is an appropriate choice in the
present problem. Clearly, $S_{4}$ is superior to $S_{2}$ in
accuracy. Thus, it will be employed in the later computations.

In fact, Orbit 1 is regular and Orbit 2 is chaotic. The regularity
or chaoticity is seen from the Poincar\'{e} map  in the
two-dimensional $r-p_r$ plane in Figure 1 (c). All points in the
two-dimensional plane are  intersections of the particles'
trajectories with the surface of section $\theta=\pi/2$ with
$p_\theta>0$ in phase space. The regular dynamics of Orbit 1 is
shown because the points in the surface of section form a closed
curve, which corresponds to a cross section of a torus in the
phase space. The chaotic dynamics of Orbit 2 is known by the
points that are distributed randomly in the Poincar\'{e} map. The
Poincar\'{e} map method is a good description of the regular and
chaotical dynamical structure of trajectories in a conservative
four-dimensional phase space.  The algorithms $S_{2}$ and $S_{4}$
exhibit good long-term performance in the conservation of the
Hamiltonian (14) or (18), irrespective of whether or not the
integrated orbit is chaotic.

\subsection{Orbital dynamics}

In addition to the Poincar\'{e} map method, Lyapunov exponents and
fast Lyapunov indicators (FLIs) are common methods to detect chaos
from order. The maximal Lyapunov exponent corresponding to the
most unstable direction in the phase space directly measures  how
the given orbit is sensitive dependence on the initial conditions.
It is defined in [51] as an invariant indicator of chaos under
spacetime coordinate transformations within the relativistic
framework
\begin{eqnarray}
    \lambda =\lim_{\tau \to \infty} \ln\frac{1}{\tau} \frac{d(\tau)}{d(0)},
\end{eqnarray}
where $d(\tau)$ and $d(0)$ are the proper distances of two nearby
orbits at the proper time $\tau$ and the starting time,
respectively. $\lambda$ tends to zero for a bounded, regular
orbit, while it attains a positive value for a bounded, chaotic
orbit. The distinct evolution tendency of $\lambda$ can
distinguish between chaotic and regular orbits. Based on this
point, the regular dynamics of Orbit 1 and the chaotic dynamics of
Orbit 2 are confirmed by the maximal Lyapunov exponents in Figure
2 (a). The FLI is a quicker method to detect chaos and
quasiperiodicity than the largest Lyapunov exponent. Its invariant
definition  [52] is
\begin{eqnarray}
    FLI=\log _{10}\frac{d(\tau)}{d(0)}.
\end{eqnarray}
If  a bounded orbit has an exponentially increasing FLI with time
$\log _{10}\tau$, it is chaotic. When a bounded orbit has an
algebraically  increasing FLI, it is ordered. The different
deviation of two adjacent orbits with time is used to distinguish
between chaotic and regular dynamics. The FLIs in Figure 2 (b)
also describe different dynamical behaviors of Orbits 1 and 2.

In what follows, we use the methods of FLIs and Poincar\'{e} map
to trace the influences of the coupling constants $c_{13}$ and
$c_{14}$ on the transition from regular to chaotic dynamics. This
consideration is based on Case (ii) with $0\leq 2c_{13}<c_{14}<2$
and Case (iii) with $0\leq c_{14}<2c_{13}<2$.

In fact, the coupling constants $c_{13}$ and $c_{14}$ considered
in Figure 1 correspond to one of the  RN type black holes in  Case
(ii) with $0\leq 2c_{13}<c_{14}<2$. In this case, we continue our
numerical simulations by taking the parameters $b=9.7\times
10^{-4}$, $E=0.994$ and $L=4.6$ with the initial separation
$r=35$. Given $c_{14}=1.4$, the values of  $c_{13}=$0.05, 0.25 and
0.40 correspond to the existence of order, weak chaos and strong
chaos in Figure 3 (a)-(c). Although the orbits  seem to be
seven-island orbits for $c_{13}=$0.05 and 0.25 from the
Poincar\'{e} map, $c_{13}=$0.05 indicates the regular dynamics and
$c_{13}=$0.25 yields the chaotic dynamics from the FLIs in Figure
3 (d). These facts seem to show that the transition from regular
to chaotic dynamics occurs easily and chaos becomes stronger as
the coupling constant $c_{13}$ increases. On the other hand, an
increase of the coupling constant $c_{14}$ seems to suppress the
occurrence of chaos and to weaken the extent of chaos when the
coupling constant $c_{13}=0.3$ is given in Figure 3 (e)-(h).

Figure 4 describes the phase space structures of orbits for
several values of $c_{14}$ or  $c_{13}$ in Case (iii) with $0\leq
c_{14}<2c_{13}<2$ corresponding to non-RN black holes, where
$b=9.4\times 10^{-4}$, $E=0.994$, $L=4.8$ and $r=35$. In Figure 4
(a)-(c), $c_{14}=0.1$ and  $c_{14}=0.3$ yield regular dynamics,
but $c_{14}=0.4$ exhibits chaotic dynamics for $c_{13}=0.50$. When
$c_{14}=0.4$ is given in  Figure 4 (c)-(e), $c_{13}=0.50$, 0.54
and 0.65 correspond to strong chaos, weak chaos and order in
sequence. This fact can also be seen clearly from the FLIs in
Figure 4 (f).

Figure 5 (a) plots the dependence of the FLI on the coupling
parameter $c_{13}$ in Case (ii), where  the other parameters and
the initial separation are those of Figure 3 (d). Each FLI is
computed until  the integration time reaches  $1\times 10^{6}$. 5
is the threshold of the FLI between chaos and order. That is, all
FLIs less than 5 indicate the regularity, whereas those larger
than or equal to 5 show the chaoticity. Chaos occurs almost
everywhere for $c_{13}>0.2611$ in Figure 5 (a). On the other hand,
there is chaos almost everywhere for $c_{14}<1.4302$ with
$c_{13}=0.3$ in Figure 5 (b). The two cases in Figure 5 (a) and
(b) are considered together in Figure 6 (a) that lists a
distribution of order and chaos in the binary  parameter space
$(c_{13},c_{14})$. The region under the line
$c_{14}=0.7767c_{13}+1.1972$  is chaotic almost everywhere, while
the region over the line is almost regular.  The dynamics in Case
(iii) is unlike that in Case (ii). Chaos occurs mainly for
$c_{13}<0.5396$ with $c_{14}=0.4$ in Figure 5 (c), and so does it
for $c_{14}>0.2610$ with $c_{13}=0.5$ in Figure 5 (d). The two
cases in Figure 5 (c) and (d) can be shown together through the
distribution of order and chaos in the binary parameter space
$(c_{13},c_{14})$ in Figure 6 (b). The main chaotic region is over
the line $c_{14}=3.5155c_{13}-1.4968$, and the ordered region is
mainly below the line.

Replacing $b$, $E$ and $L$ in Figure 6 (a) with $b=9.1\times
10^{-4}$, $E=0.995$ and $L=$4.8, 5.0, 5.2, we give distributions
of order and chaos in the binary  parameter space
$(c_{13},c_{14})$ in Figure 7 (a)-(c). Based on Case (ii), the
distributions are typically different for distinct values of the
angular momentum $L$. When Case (ii) gives place to Case (iii) in
Figure 7 (d)-(f), the dependence of chaotic dynamics on $c_{13}$
or $c_{14}$ is varied with the variation of  $L$. The chaotic
region is enlarged as the angular momentum $L$ increases. The
result is unlike that of Refs. [44,45] on the increase of $L$
weakening the degree of chaos.

Figure 8 (a)-(c) relates to the distributions of order and chaos
for several different values of the magnetic parameter $b$ with
$L=4.7$ and $E=0.995$ in Case (ii). The chaotic region increases
with the magnetic parameter $b$ increasing. This result is also
suitable for the increase of energy $E$ in Figure 9 (a)-(c) and
initial radius $r$ in Figure 10 (a)-(c). The distributions of
order and chaos for several values of $b$, $E$ or $r$ in Case
(iii) (i.e., Figures 8-10 (d)-(f))  are different from those in
Case (ii).

In short, no universal rule can be given to the dependence of
regular and chaotic dynamics on varying one or two parameters
$c_{13}$ and $c_{14}$ in Cases (ii) and (iii). The distributions
of order and chaos in the binary  parameter space
$(c_{13},c_{14})$ are different when different combinations of the
other parameters $E$, $L$, $b$ and the initial radius $r$ are
considered. This result should be reasonable because not only one
or two parameters but also a combination of the other parameters
and the initial conditions are responsible for the occurrence of
chaos.

\section{Conclusion}
\label{sec1}

The Einstein-\AE ther gravity  is a covariant theory of gravity
violating the local Lorentz symmetry. In the Einstein-\AE ther
metric describing a non-rotating black hole, there are two  free
parameters $c_{13}$ and $c_{14}$. This metric corresponds to the
RN black hole solution when the two  free parameters satisfy the
condition $0\leq 2c_{13}<c_{14}<2$, but it is a  non-RN black hole
solution for the case of
 $0\leq c_{14}<2c_{13}<2$. The spacetime is integrable in the two cases.
When the black hole is immersed in an external asymptotically
uniform magnetic field, the dynamic of charged particles moving
around the non-rotating black hole is not integrable.

There are five explicitly integrable splitting terms in the
Hamiltonian system describing the motion of charged test particles
around the Einstein-\AE ther black hole surrounded by the external
magnetic field. This gives a good chance for the construction of
explicit symplectic integrators. Numerical tests show that the
established second- and fourth-order explicit symplectic methods
exhibit excellent long-term performance in accuracy of conserving
the Hamiltonian. The fourth-order explicit symplectic scheme is
chosen to investigate the dynamics of charged particles.

There is no universal rule for the dependence of regular and
chaotic dynamics on varying one or two parameters $c_{13}$ and
$c_{14}$ in the two cases of $0\leq 2c_{13}<c_{14}<2$ and  $0\leq
c_{14}<2c_{13}<2$. The distributions of order and chaos in the
binary  parameter space $(c_{13},c_{14})$ rely on different
combination of the other parameters and the initial conditions.

\textbf{Author Contributions}: C.L. made contributions to the
software, writing --- original draft, and methodology. X.W.
contributed to the supervision, conceptualization, writing
--- review and editing, and funding acquisition. All authors have
thoroughly reviewed and approved the published version of the
manuscript.

\textbf{Funding}: This research has been supported by the National
Natural Science Foundation of China (Grant No. 11973020) and the
National Natural Science Foundation of Guangxi (No.
2019GXNSFDA245019).

\textbf{Data Availability Statement}: The study does not report
any data.

\textbf{Conflicts of Interest}: The authors declare no conflict of
interest.


\begin{figure*}[htpb]
        \centering{
        \includegraphics[width=10pc]{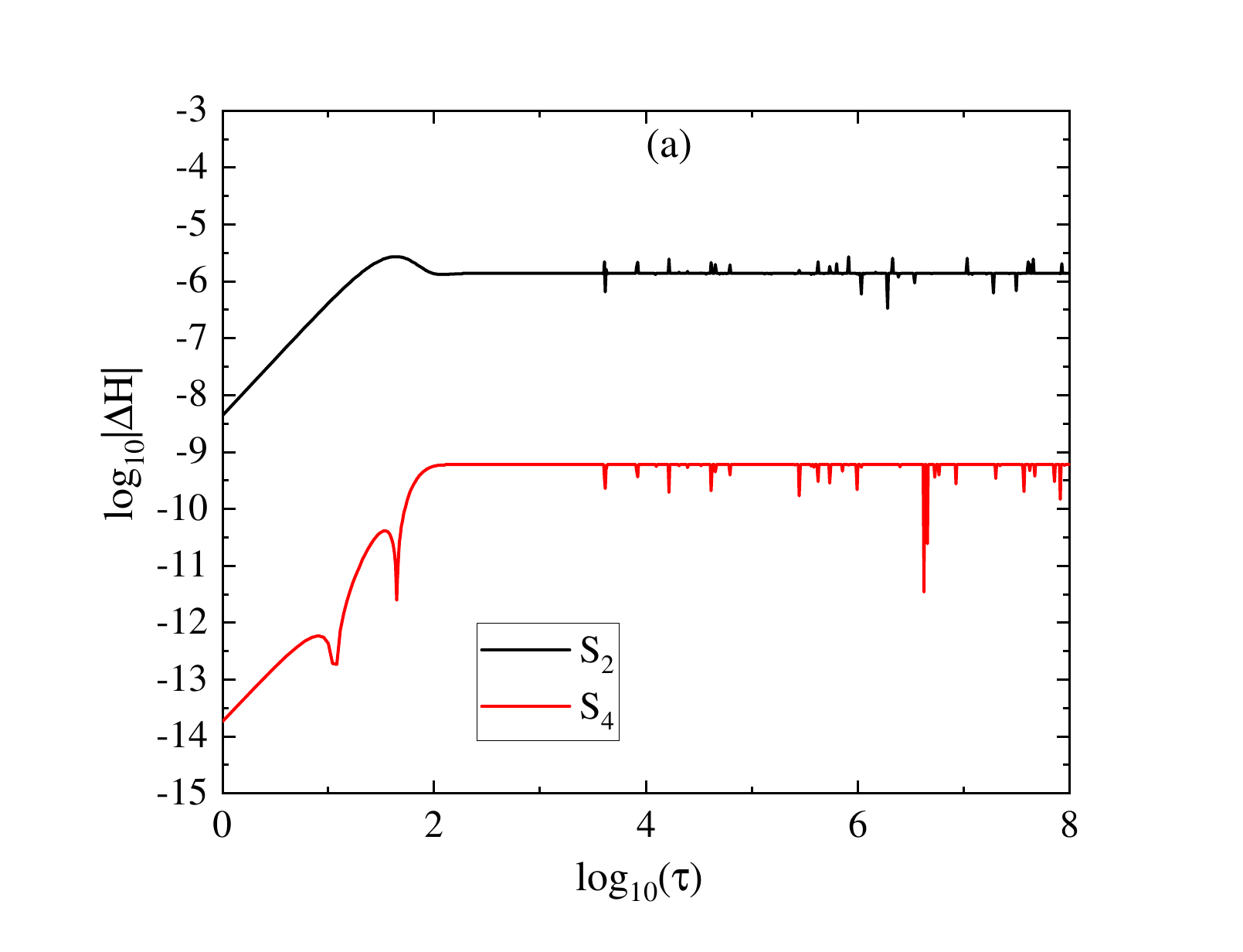}
        \includegraphics[width=10pc]{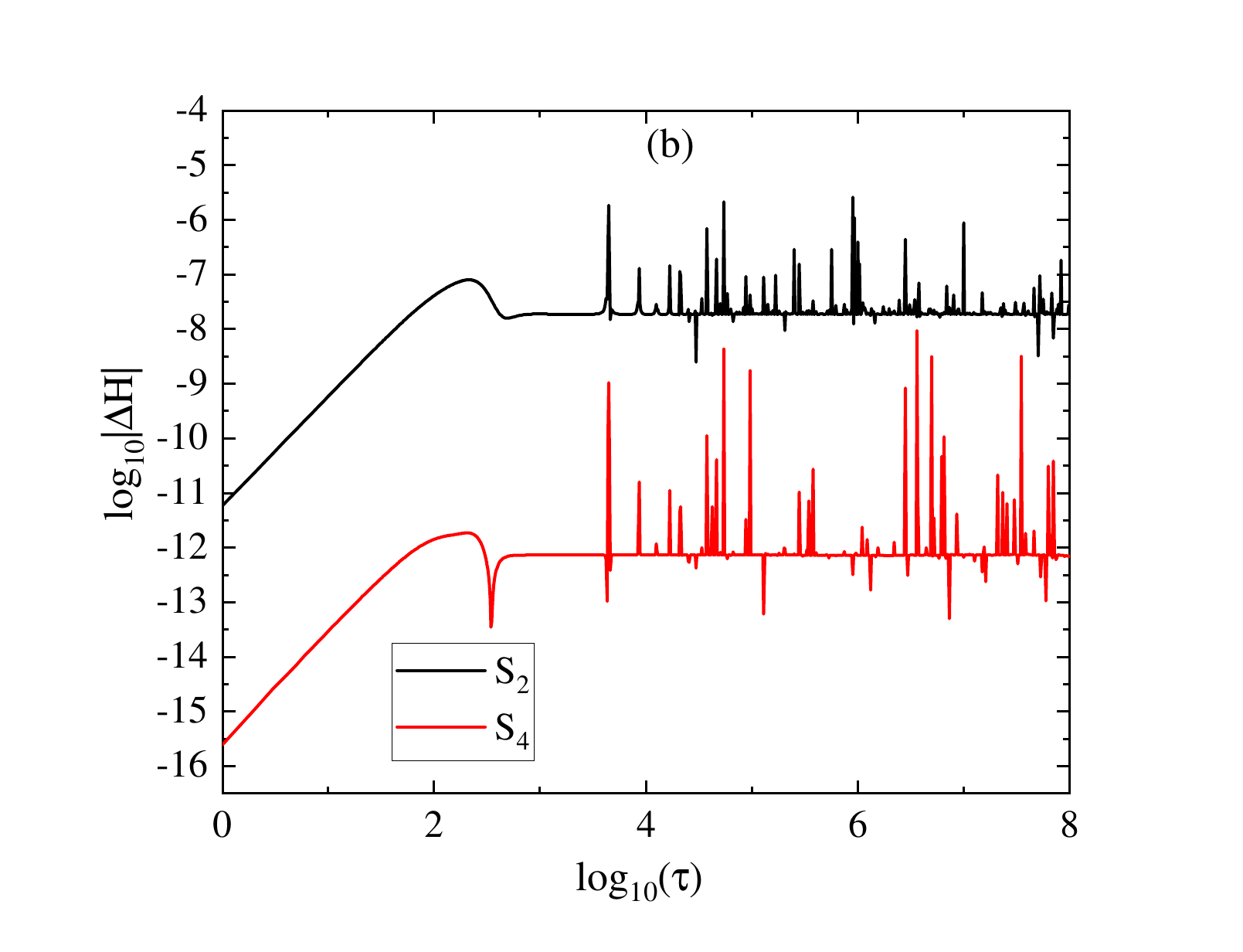}
        \includegraphics[width=10pc]{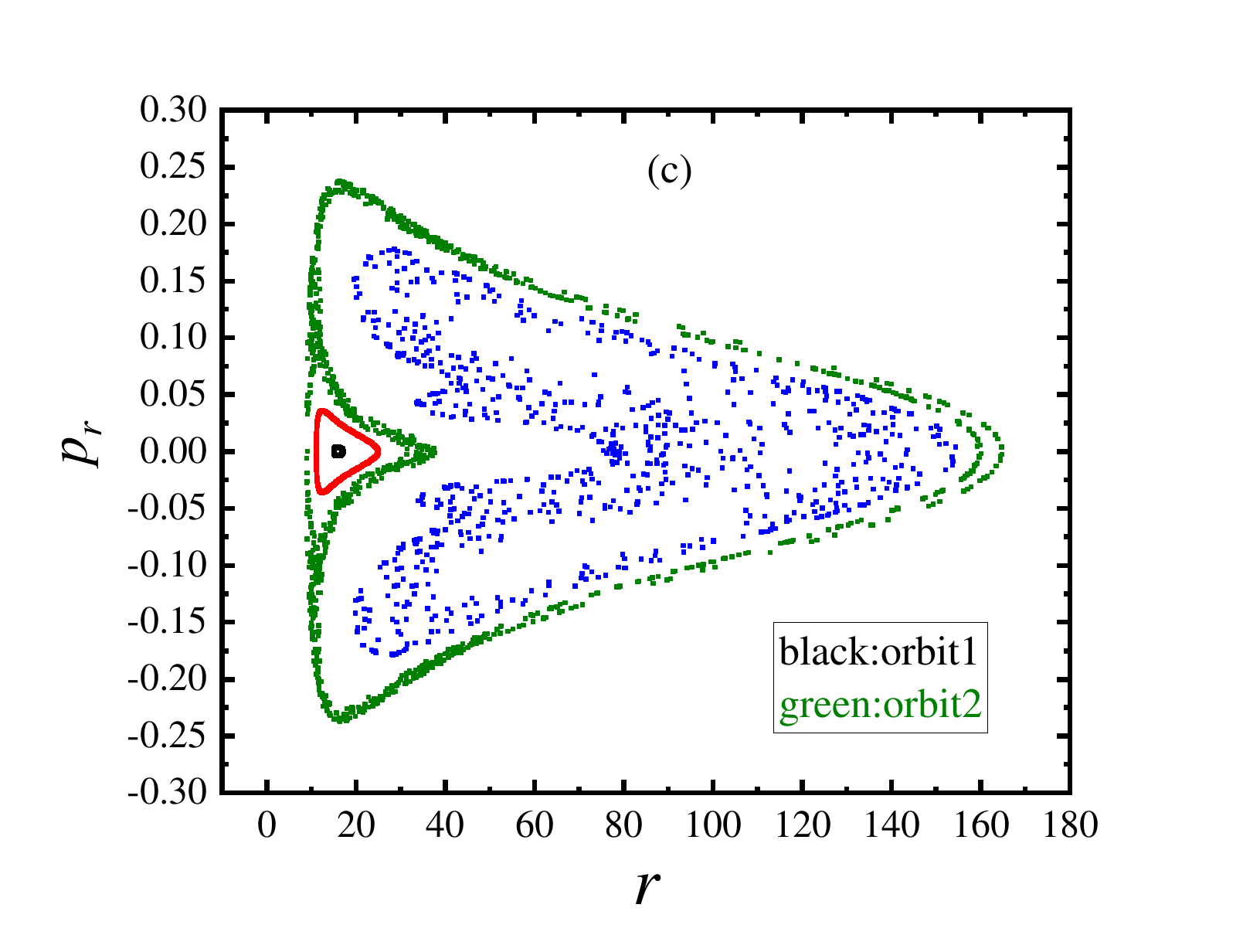}
        \caption{  (a) Hamiltonian errors $\Delta H=H+1/2$ in Equation (17) for the two
            sympletic methods $S_1$ and $S_{4}$ acting on Orbit 1.
            (b) Same as (a) but Orbit 1 replaced with Orbit 2.
            (c) Poincar\'{e} map at the plane $\theta=\pi /2$ with
             $p_{\theta}>0$. The
            parameters are $E=0.995$, $L=4.7$, $b=9.1\times
            10^{-4}$, $q=1$, $c_{13}=0.01$ and $c_{14}=0.03$ in Case (ii); the initial conditions are $p_{r}=0$
            and $\theta =\pi /2$. Orbit 1 with the initial separation $r=15$
            is regular, whereas Orbit 2 with the initial separation
            $r=35$ is chaotic.
        }
    }
\end{figure*}

\begin{figure*}[htpb]
    \centering{
        \includegraphics[width=15pc]{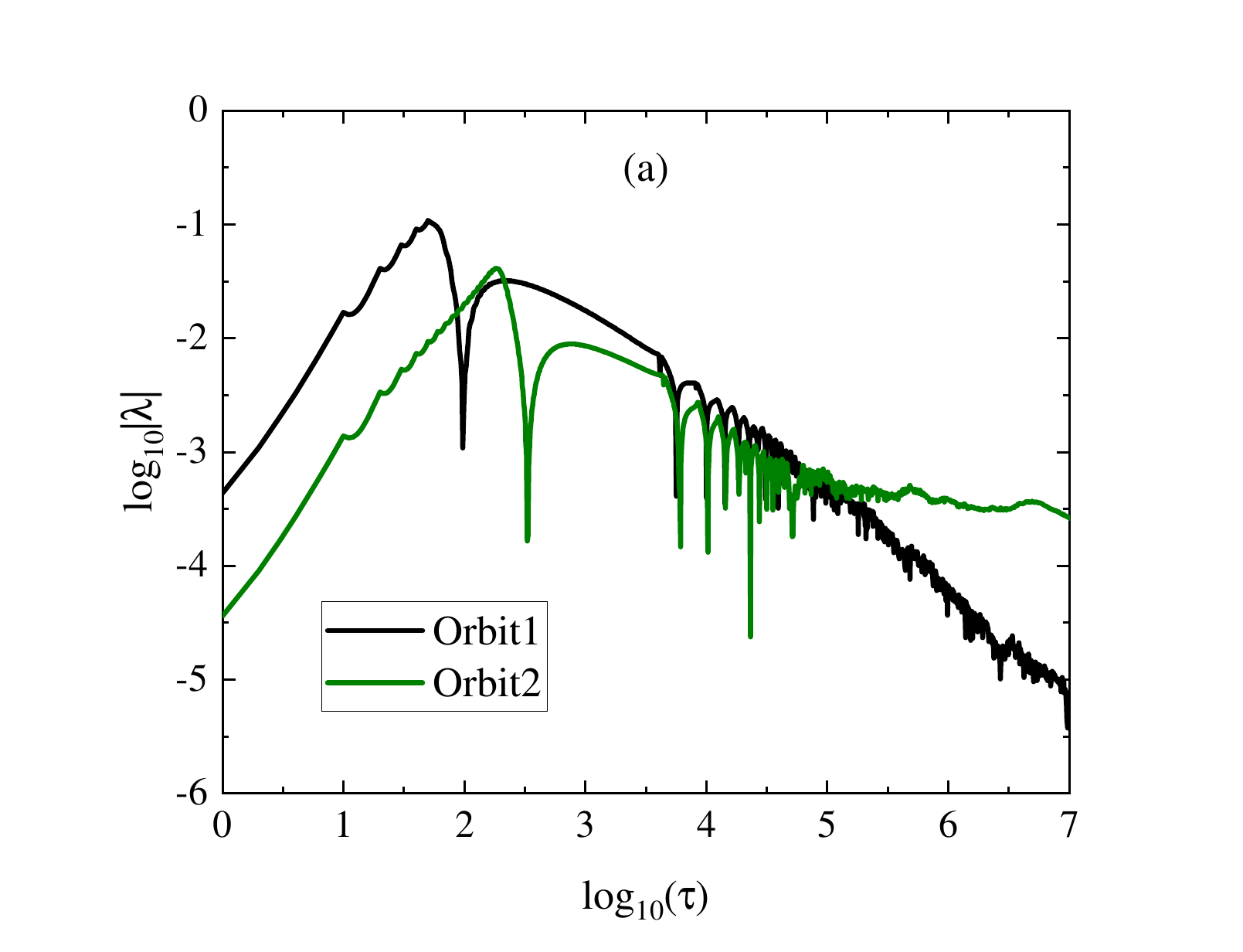}
        \includegraphics[width=15pc]{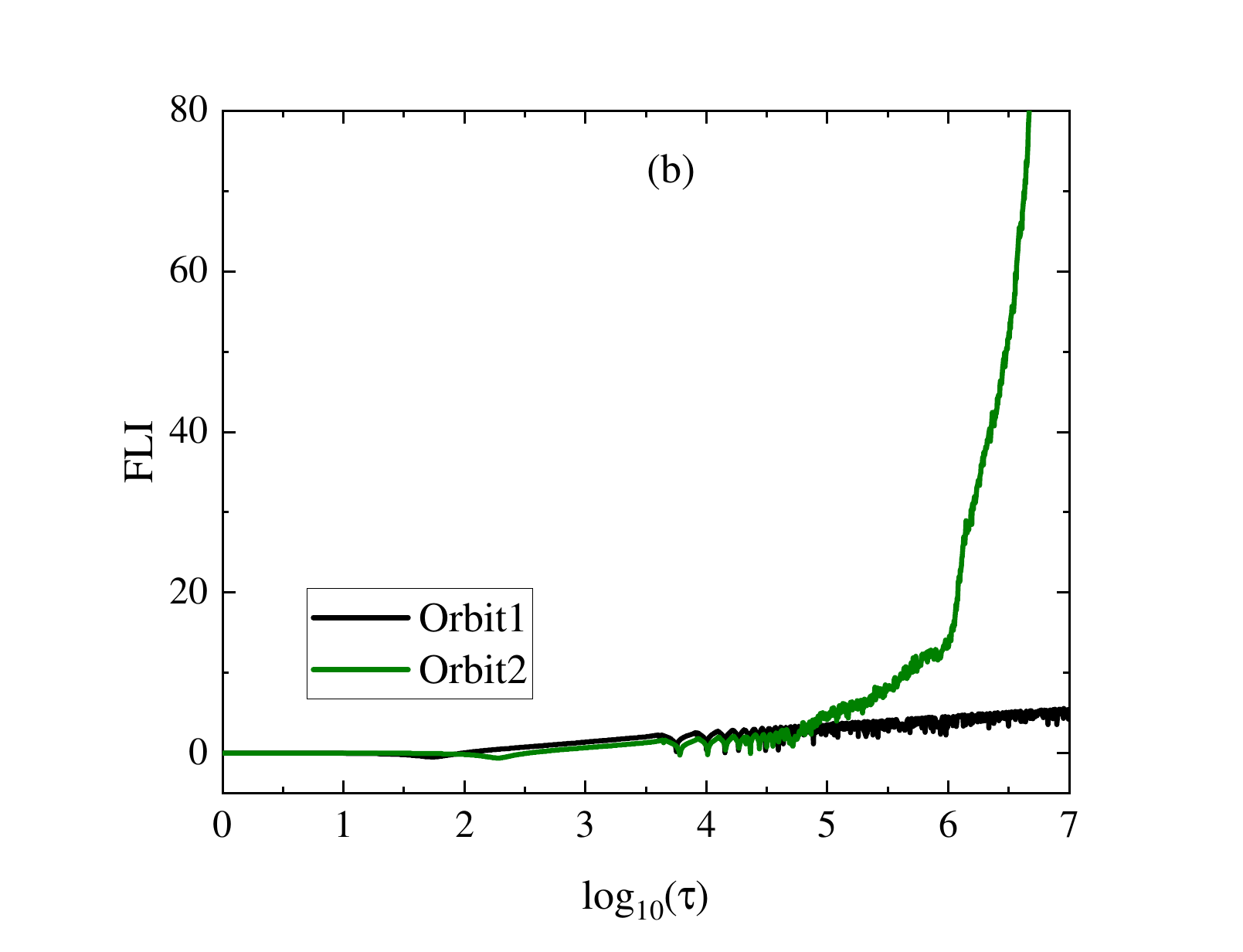}
        \caption{(a) The largest Lyapunov exponents $\lambda$
            for Orbits 1 and 2.
            (b) The fast Lyapunov indicators (FLIs) for Orbits 1 and 2.
        }
    }
\end{figure*}

\begin{figure*}[htpb]
        \centering{
        \includegraphics[width=8pc]{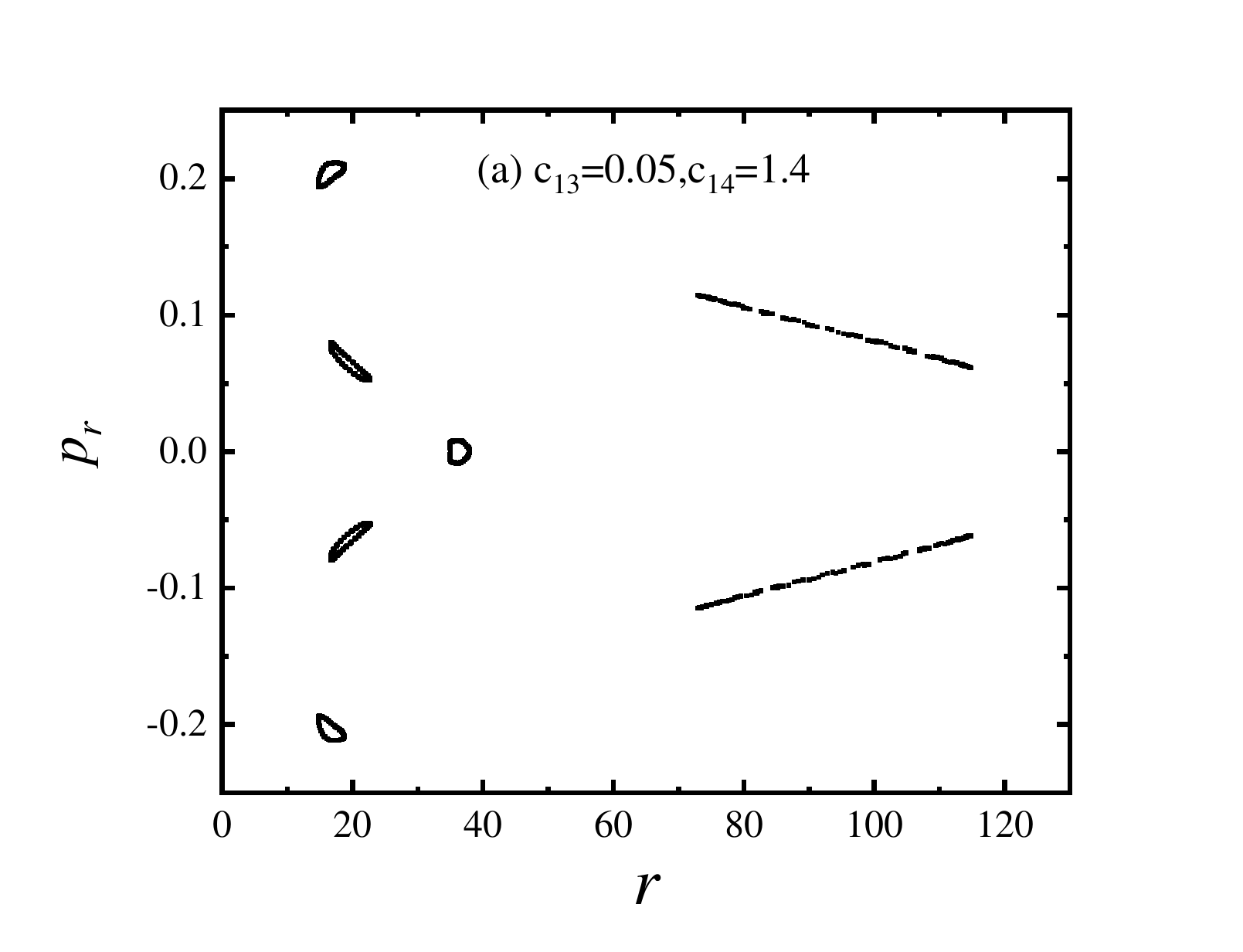}
        \includegraphics[width=8pc]{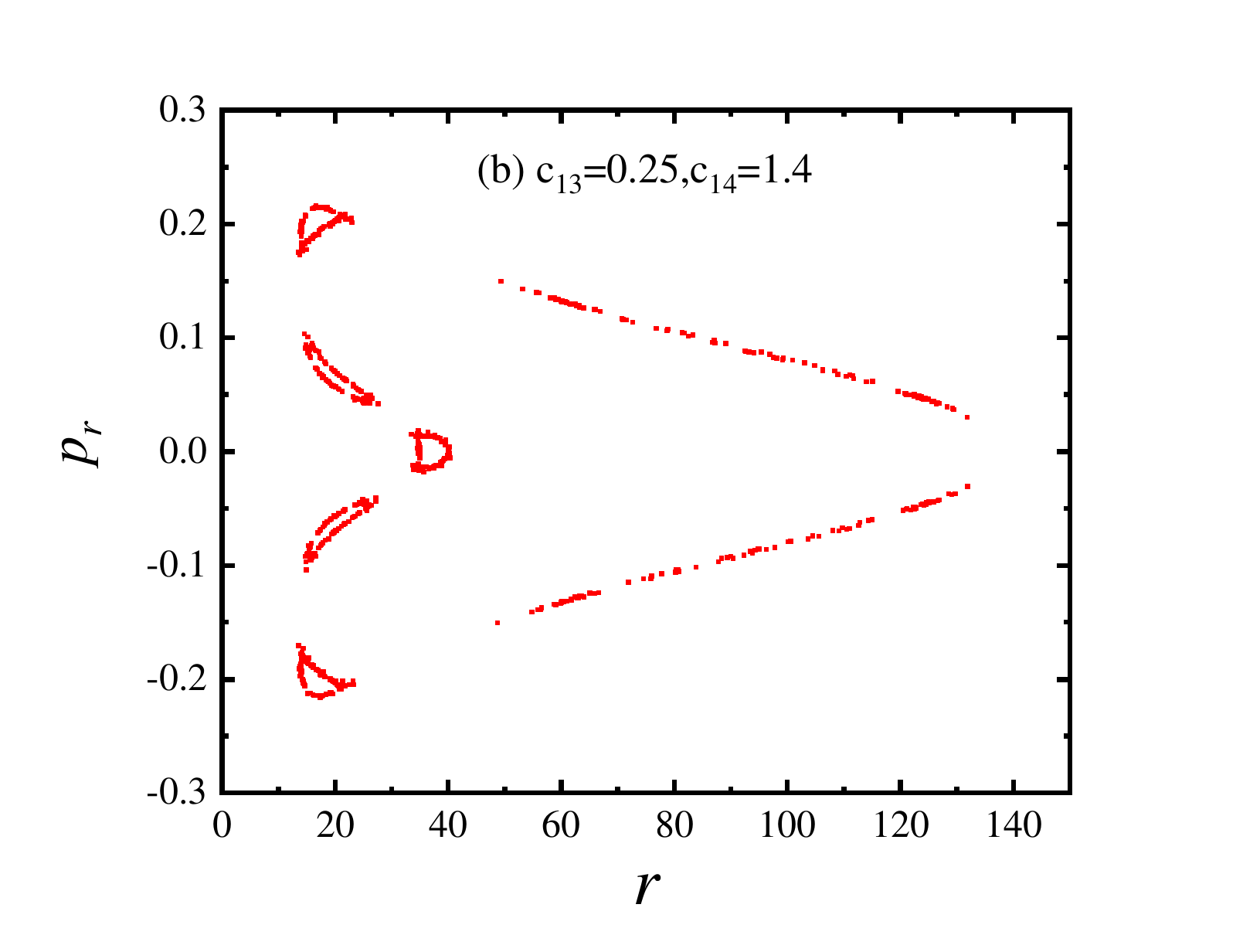}
        \includegraphics[width=8pc]{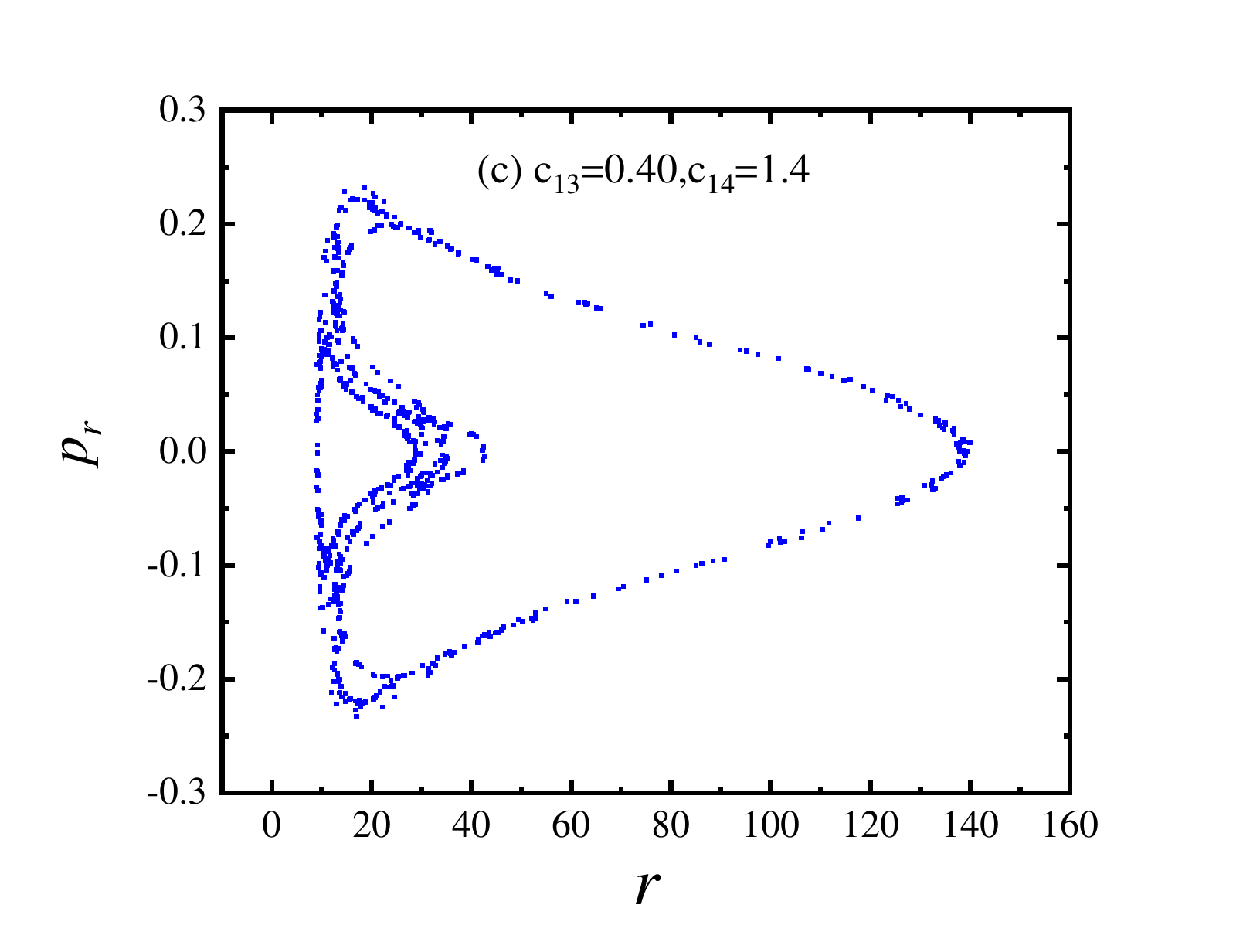}
        \includegraphics[width=8pc]{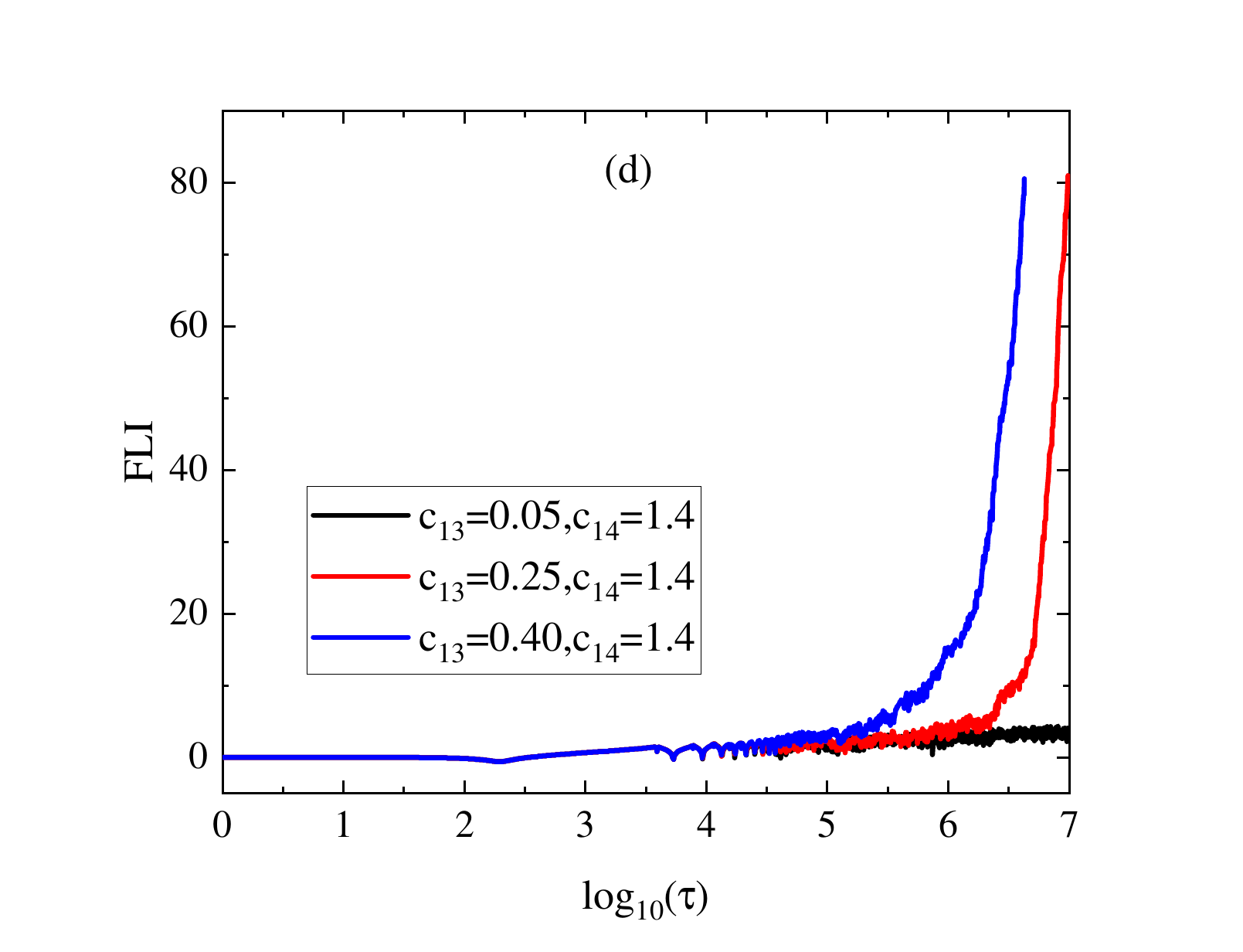}
        \includegraphics[width=8pc]{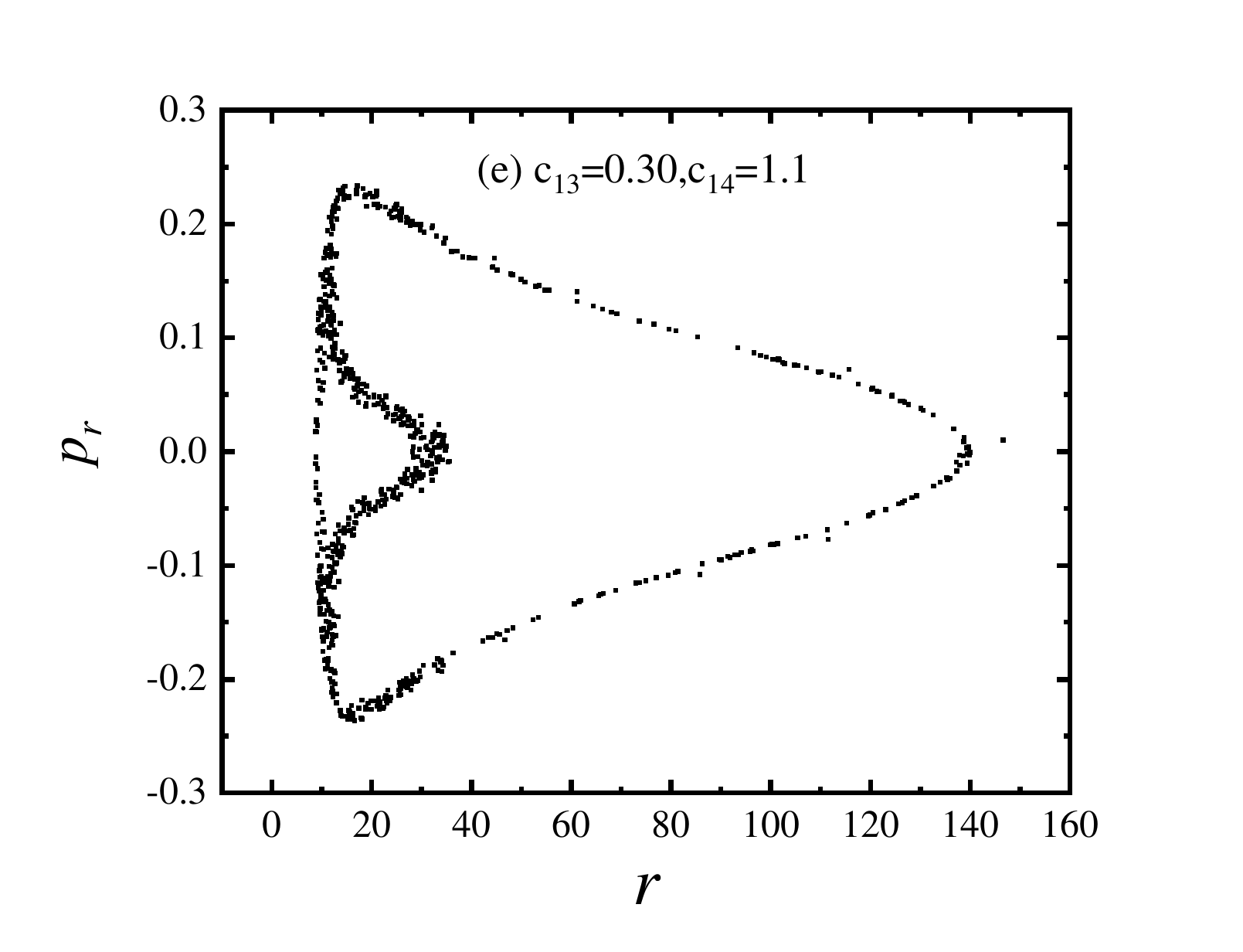}
        \includegraphics[width=8pc]{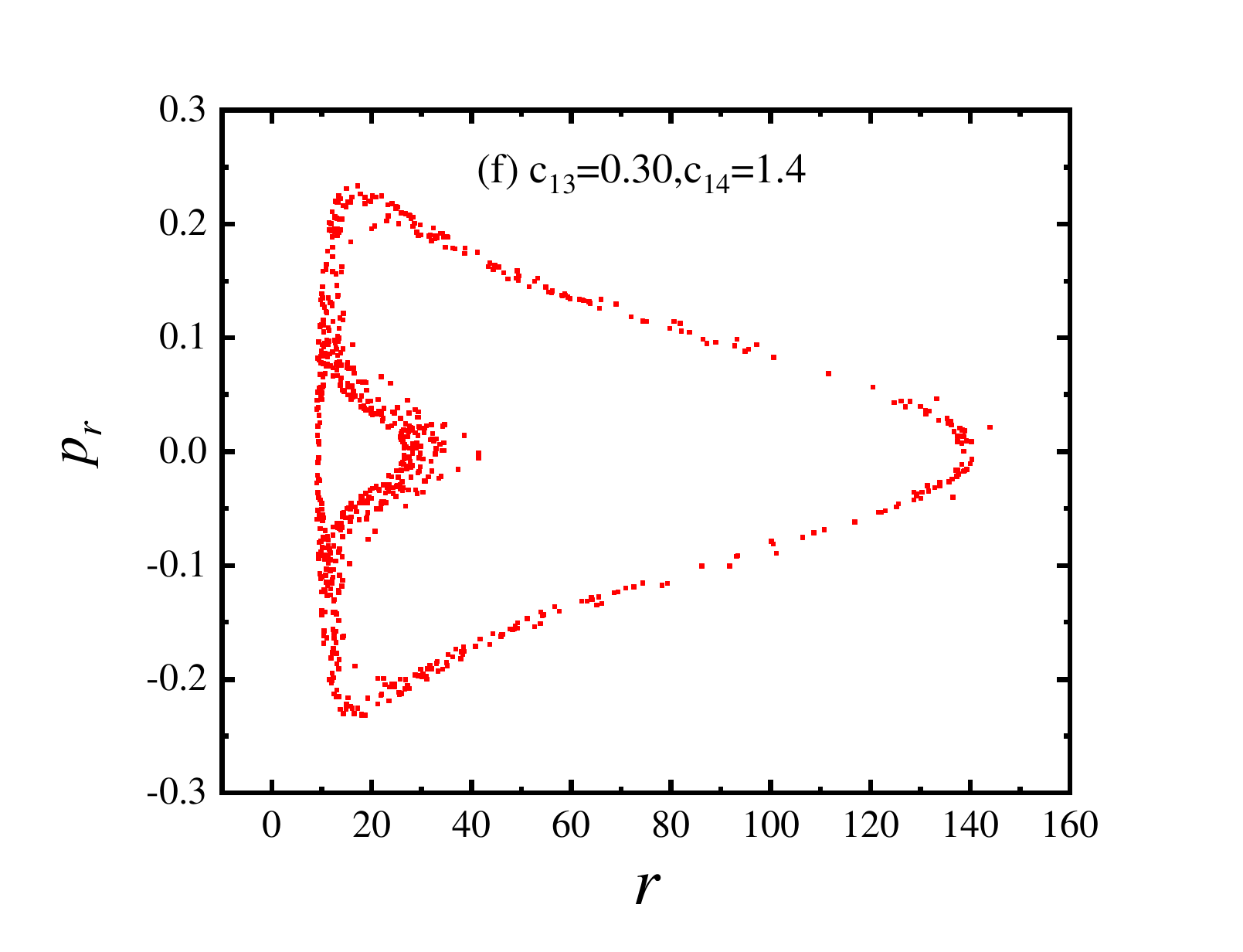}
        \includegraphics[width=8pc]{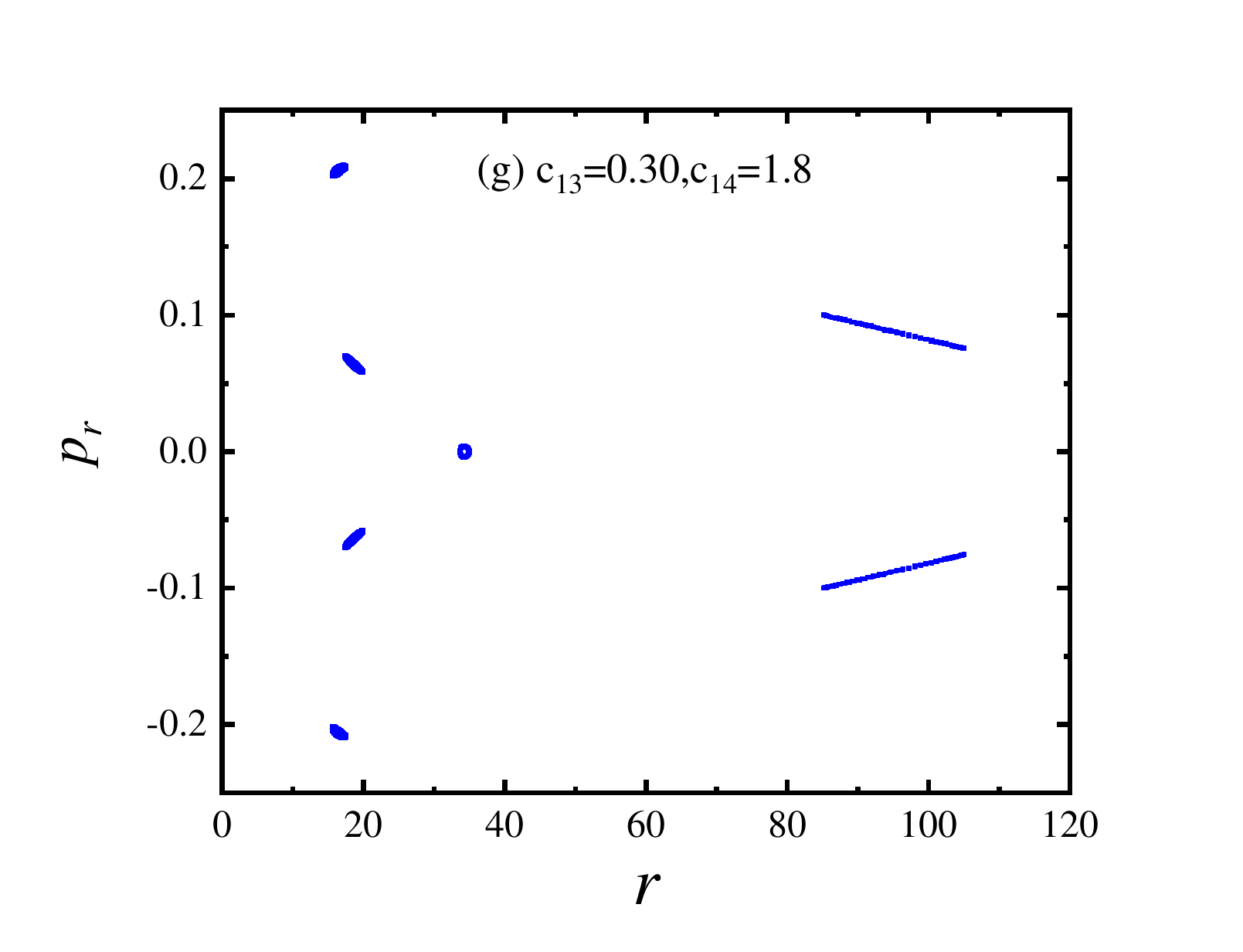}
        \includegraphics[width=8pc]{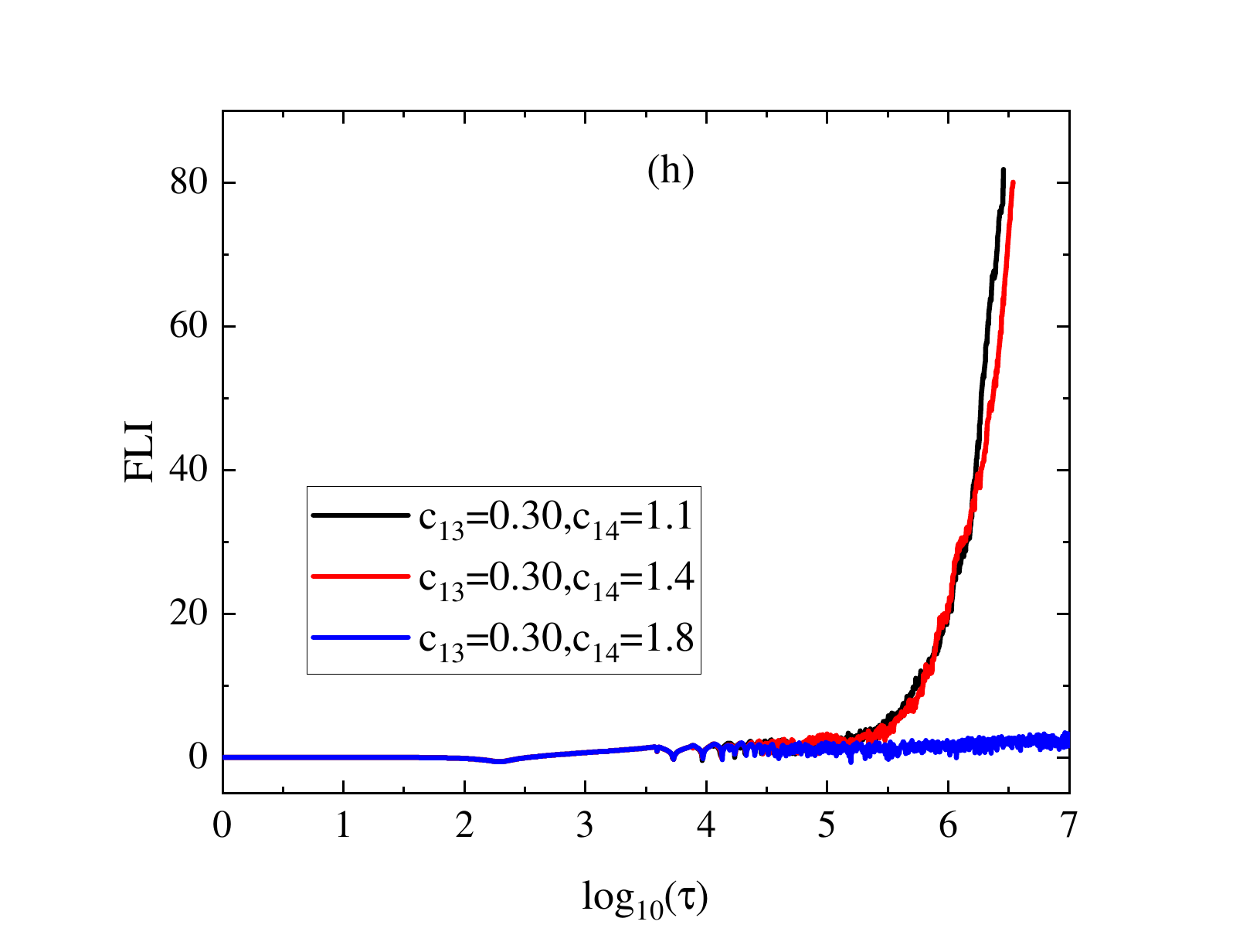}
        \caption{(a)-(c): Poincar\'{e} sections in Case (ii) of $0\leq2c_{13}<c_{14}<2$. The parameters are $b=9.7\times 10^{-4}$, $E=0.994$,
$L=4.6$, $c_{14}=1.4$, and $c_{13}$ is given different values in
(a)-(c). The initial radius is $r=35$. (d): The FLIs for the
orbits in (a)-(c). (e)-(h): Same as (a)-(d), but $c_{13}=0.30$ is
given and $c_{14}$ has distinct values.
        }
    }
\end{figure*}

\begin{figure*}[htpb]
        \centering{
        \includegraphics[width=10pc]{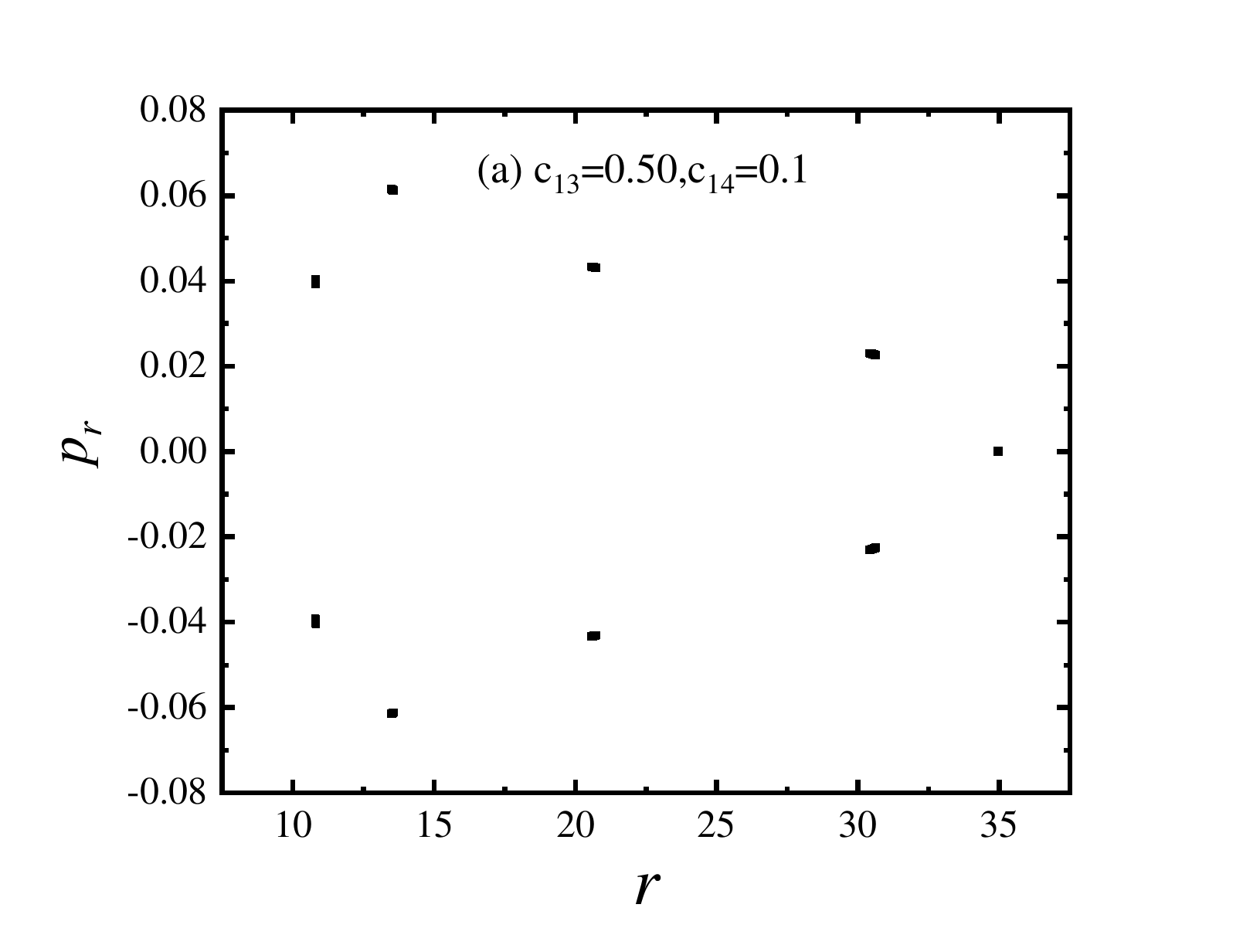}
        \includegraphics[width=10pc]{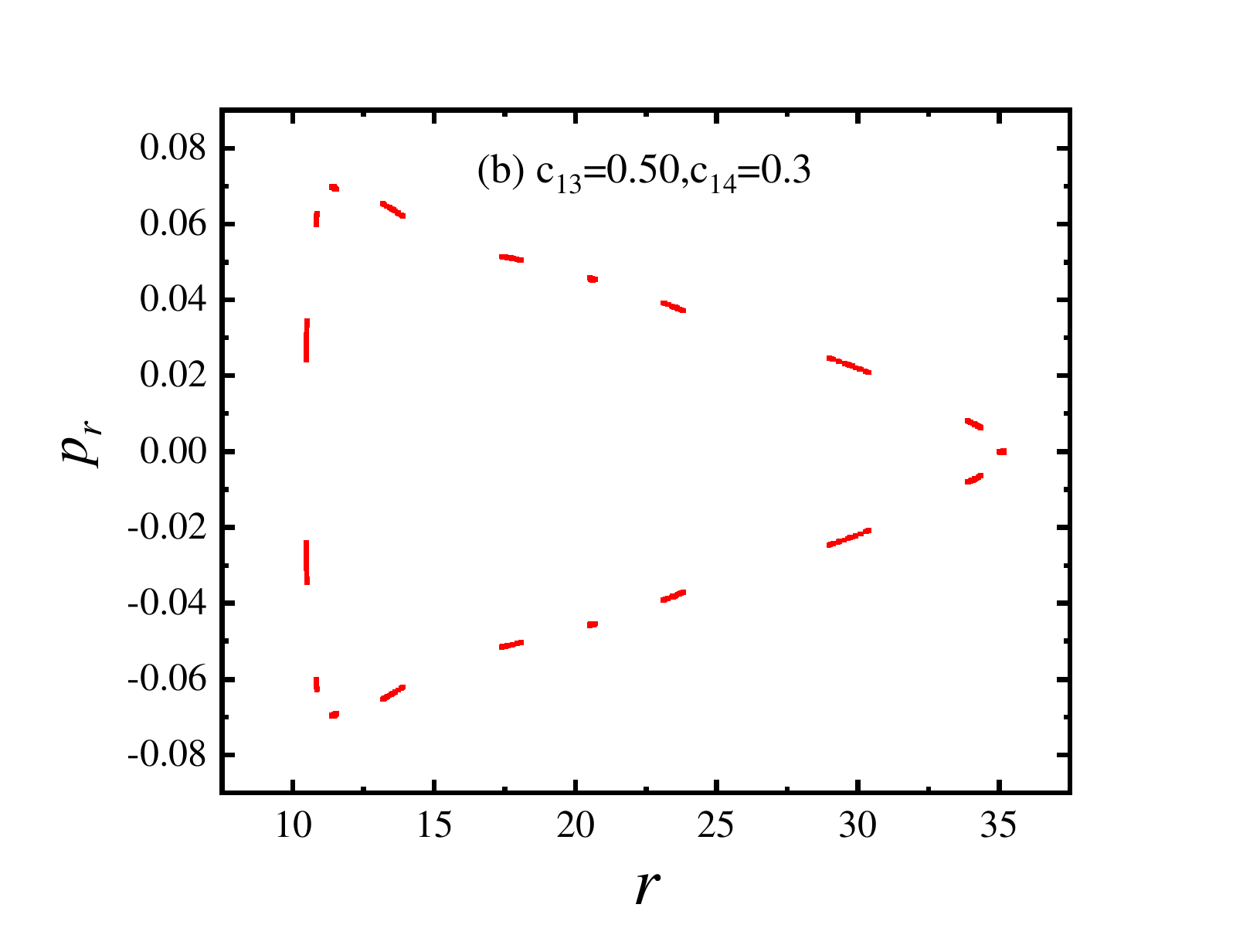}
        \includegraphics[width=10pc]{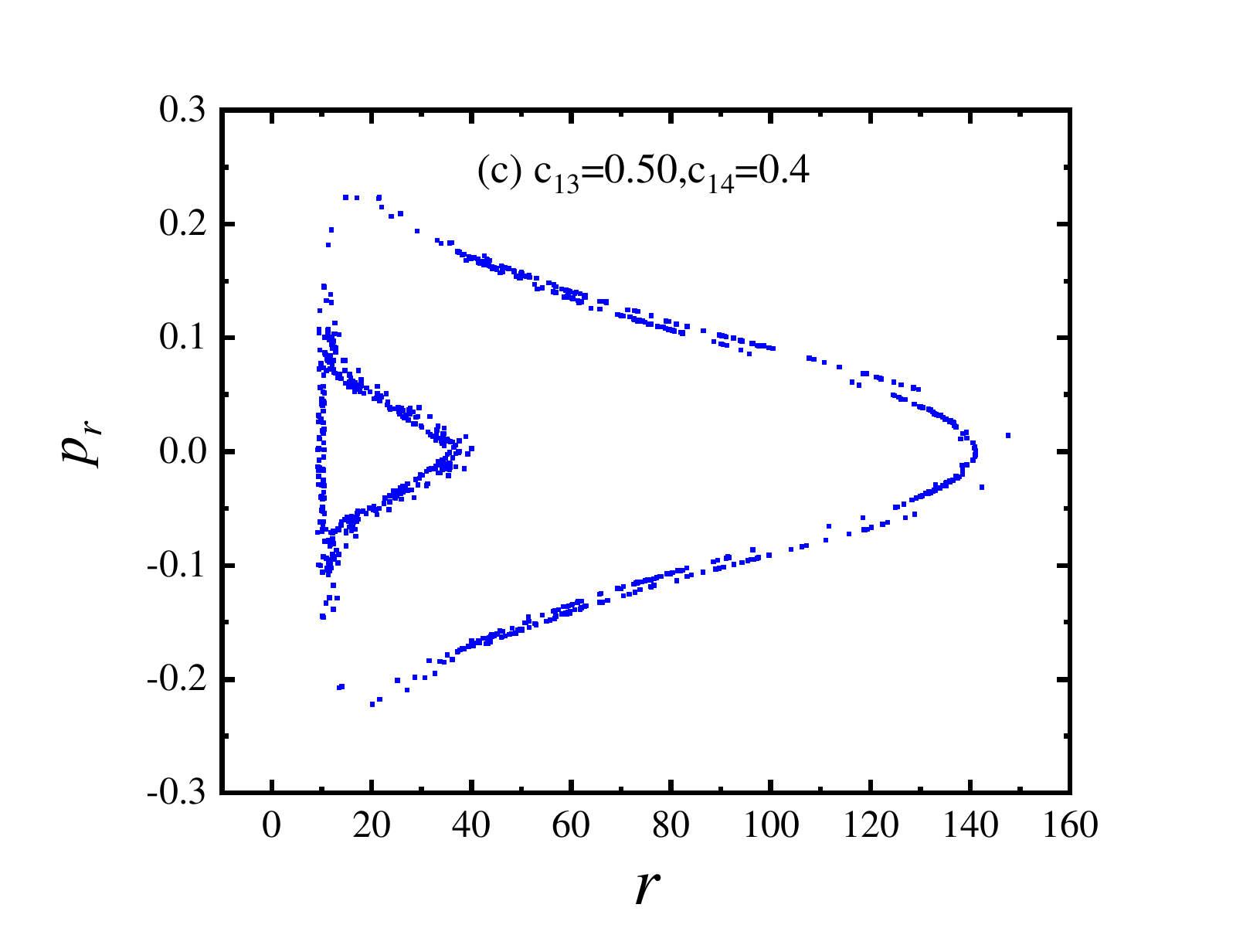}
        \includegraphics[width=10pc]{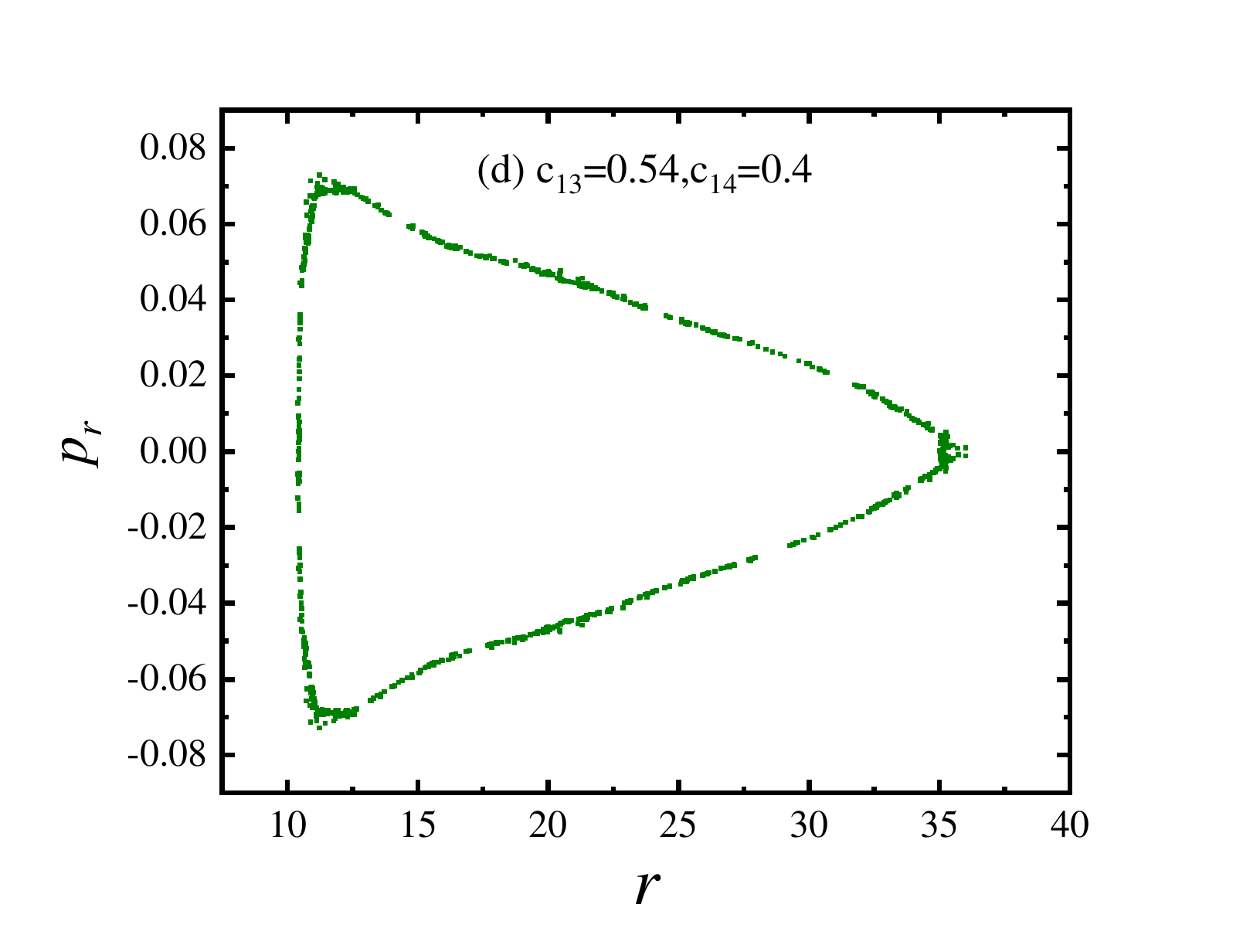}
        \includegraphics[width=10pc]{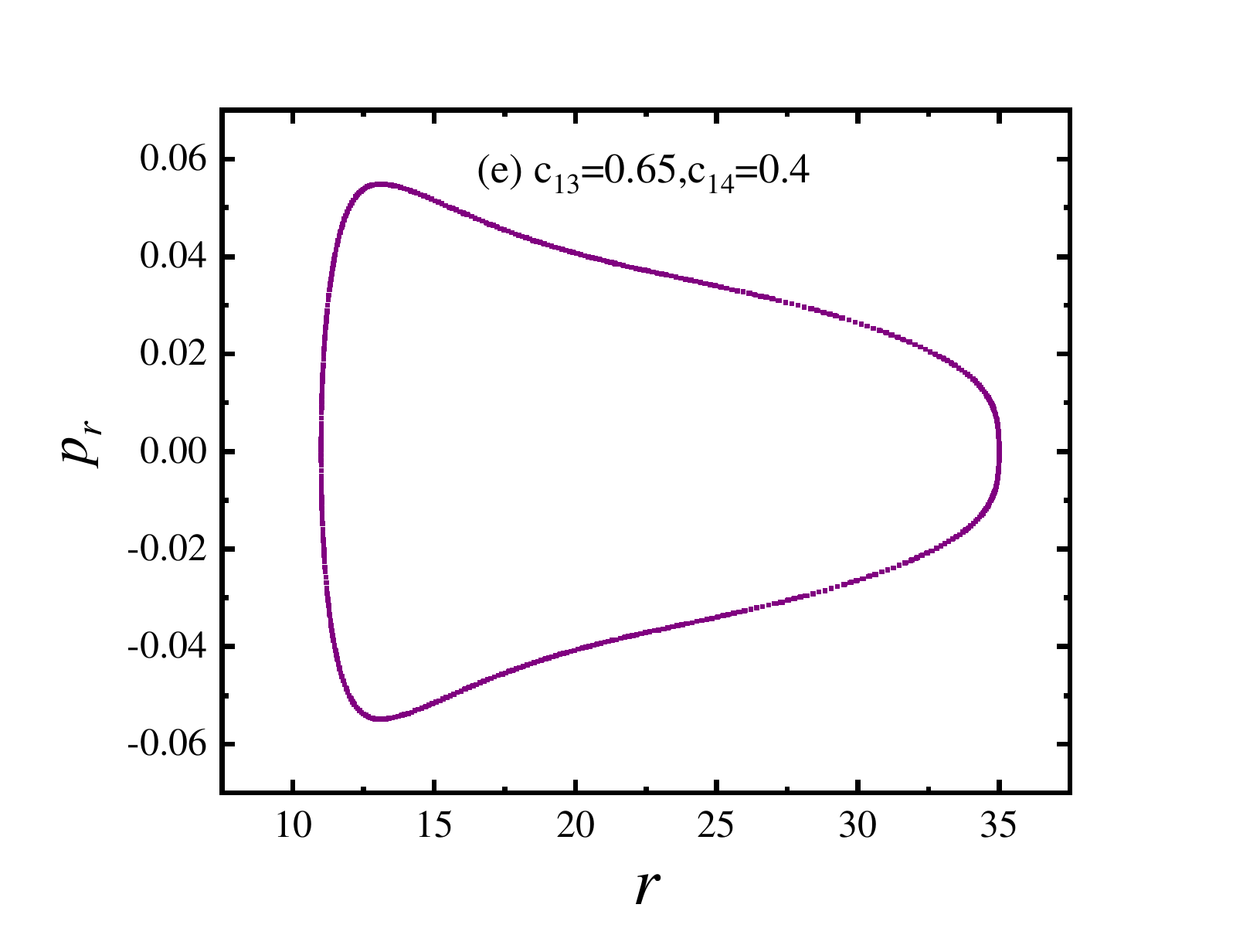}
        \includegraphics[width=10pc]{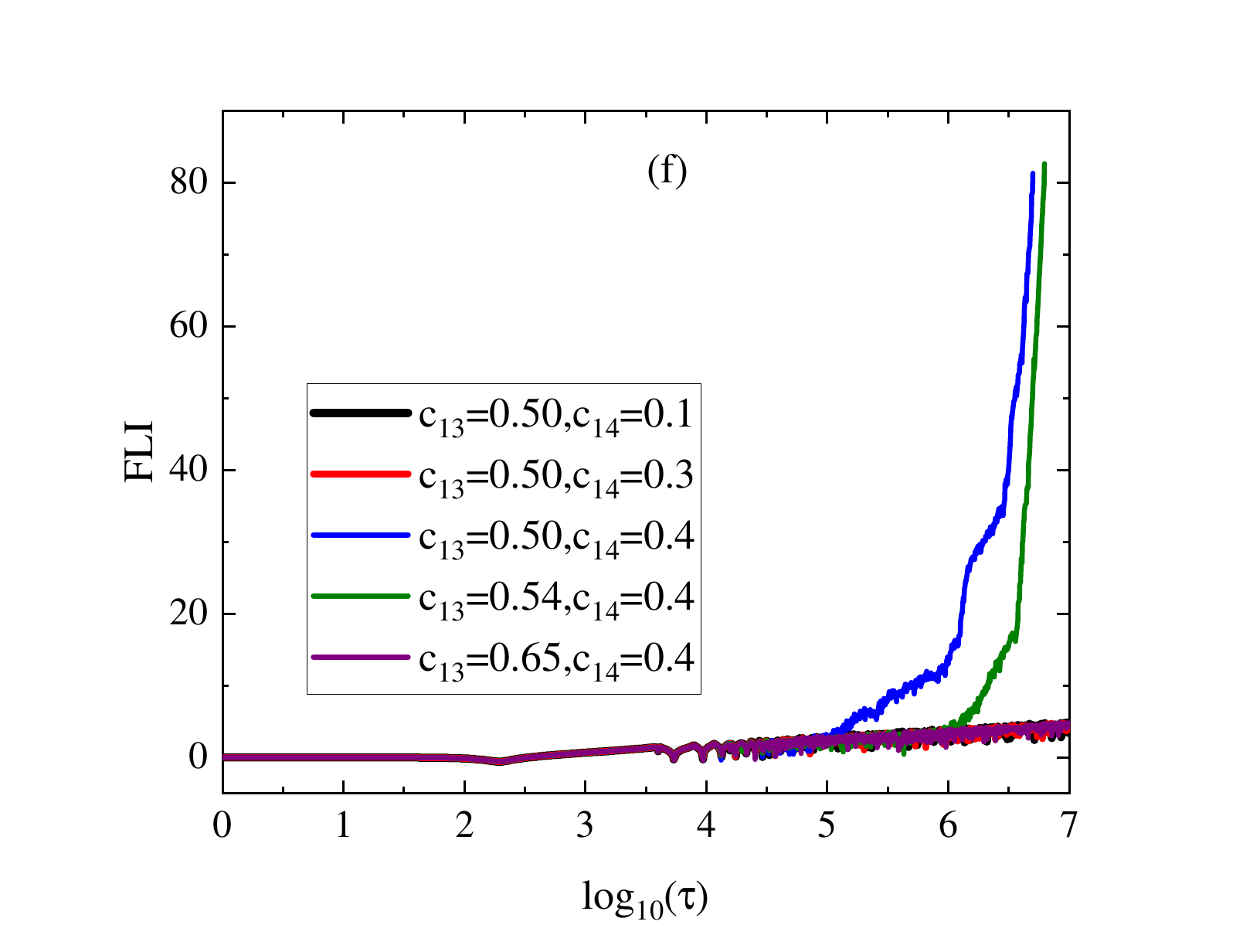}
\caption{(a)-(e): Poincar\'{e} sections in Case (iii) of $0\leq
c_{14}<2c_{13}<2$. The parameters are $b=9.4\times 10^{-4}$,
$E=0.994$, $L=4.8$; the initial radius is $r=35$. In (a)-(c),
$c_{13}=0.50$, and $c_{14}$ has different values. In (d) and (e),
$c_{14}=0.40$, and $c_{13}$ has two different values. (f): The
FLIs for the orbits in (a)-(e).
                    }
    }
\end{figure*}

\begin{figure*}[htpb]
    \centering{
        \includegraphics[width=15pc]{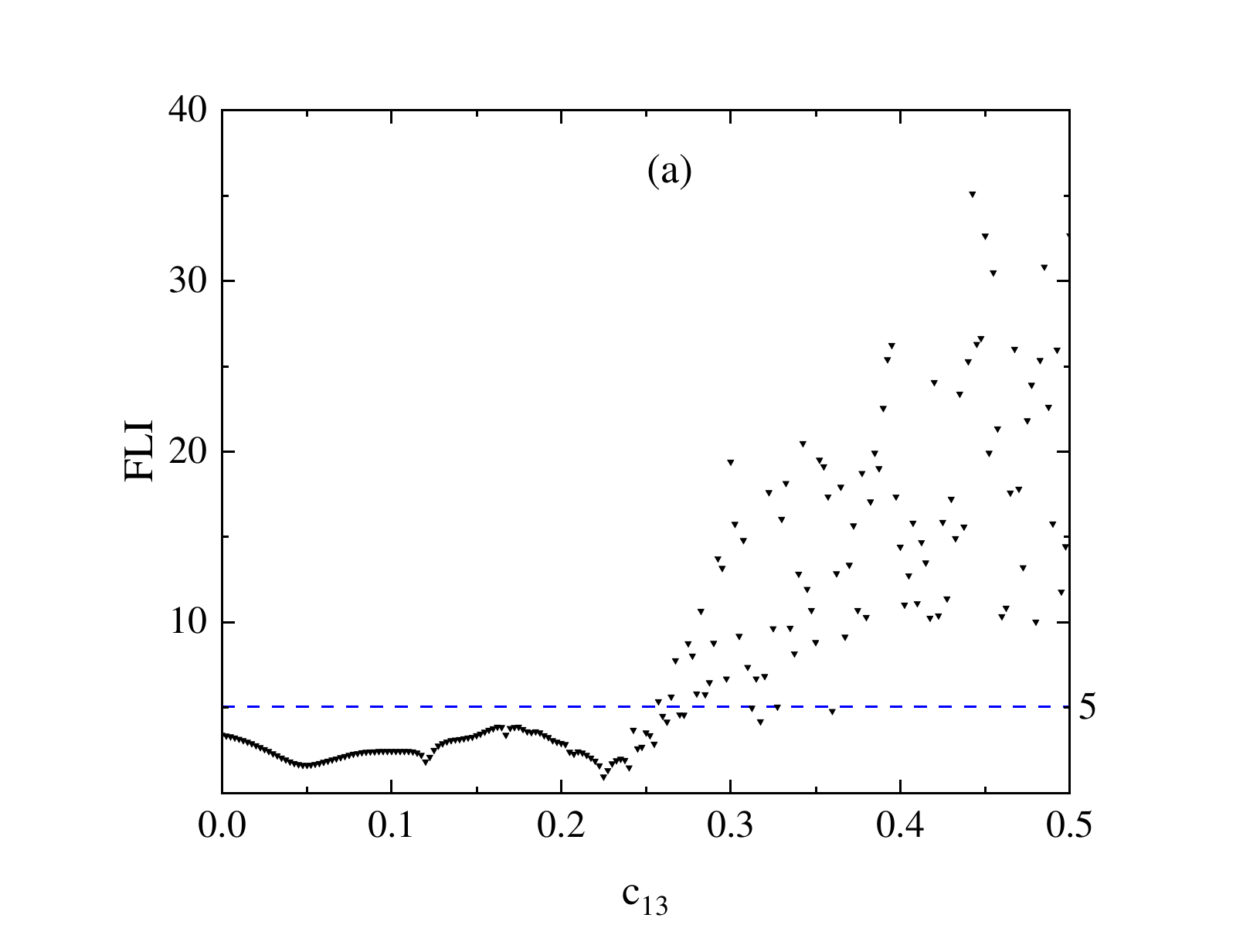}
        \includegraphics[width=15pc]{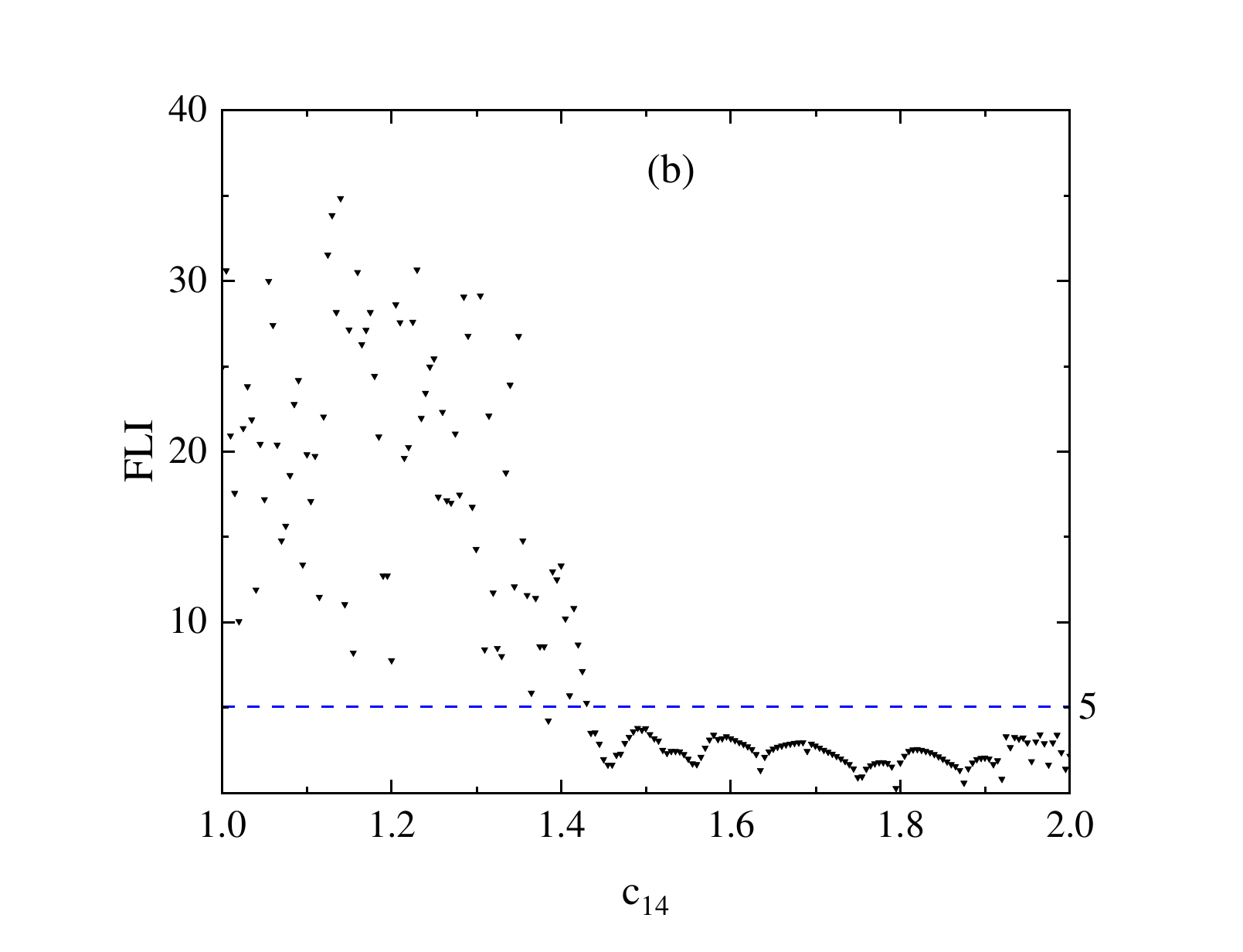}
        \includegraphics[width=15pc]{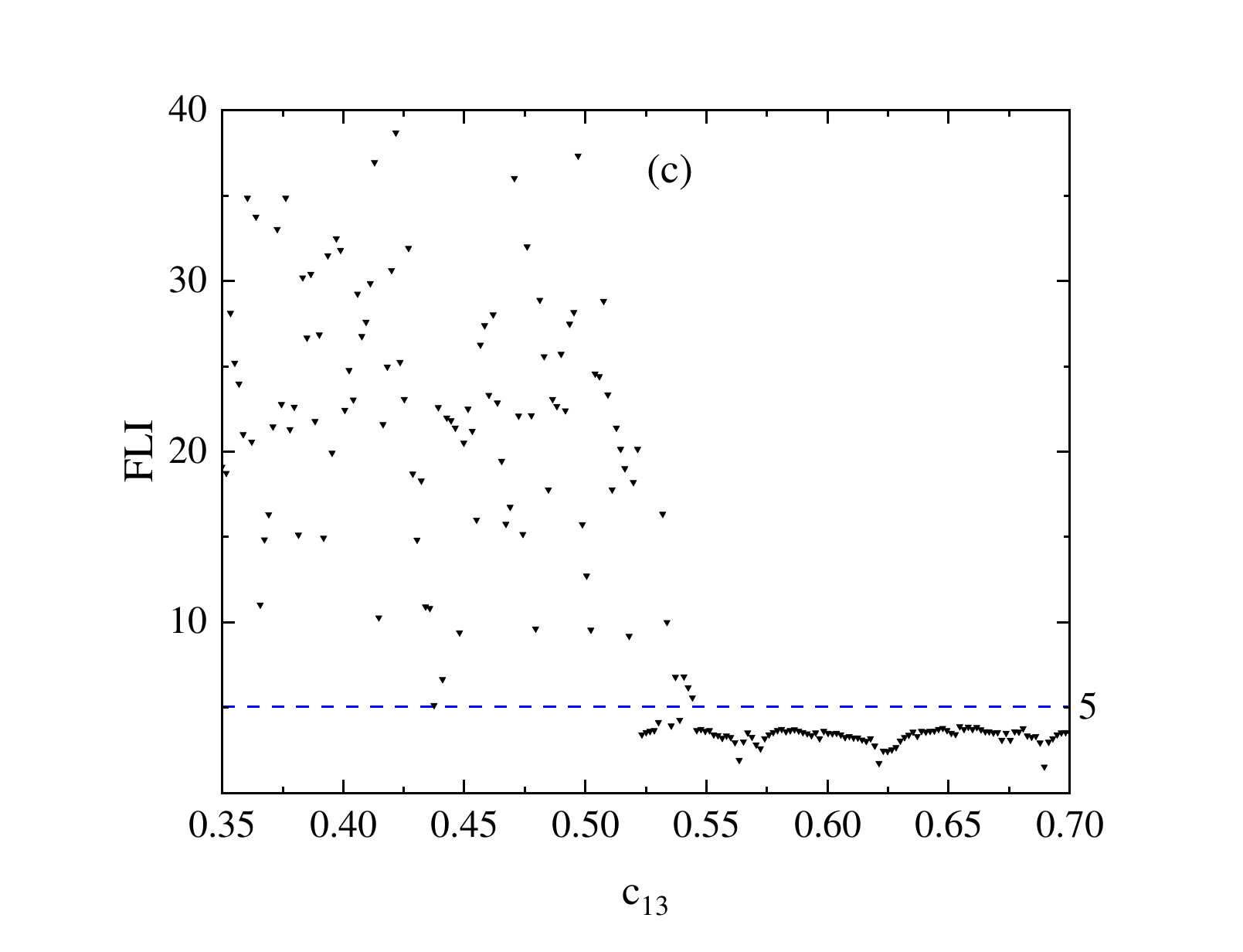}
        \includegraphics[width=15pc]{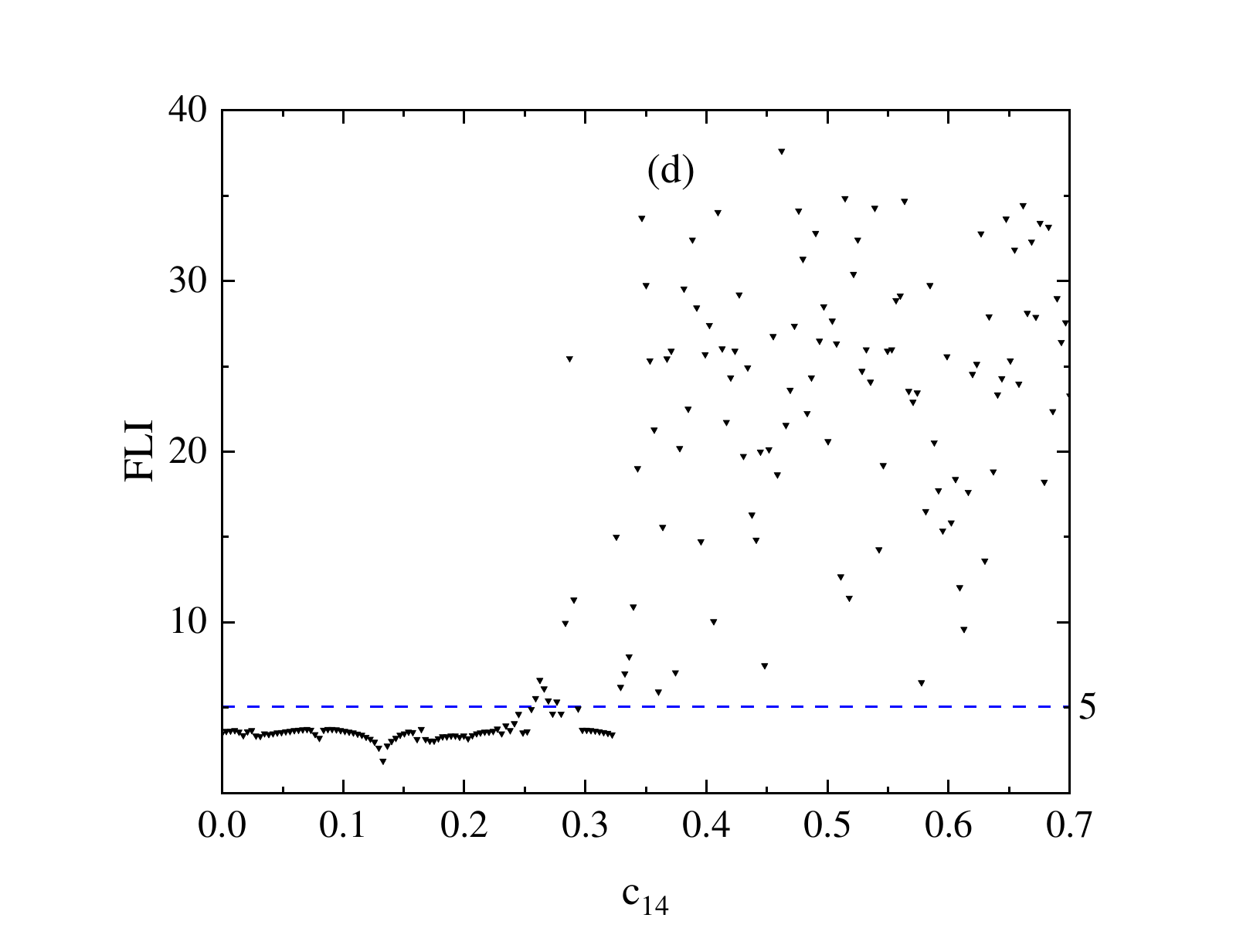}
\caption{Dependence of FLI on $c_{13}$ or $c_{14}$. (a) and (b):
Corresponding to Figure 3. $c_{14}=1.4$ is fixed in (a), and
$c_{13}=0.3$ is fixed in (b). (c) and (d): Corresponding to Figure
4. $c_{14}=0.4$ is fixed in (c), and  $c_{13}=0.5$ is fixed in
(d).
        }
    }
\end{figure*}

\begin{figure*}[htpb]
    \centering{
        \includegraphics[width=15pc]{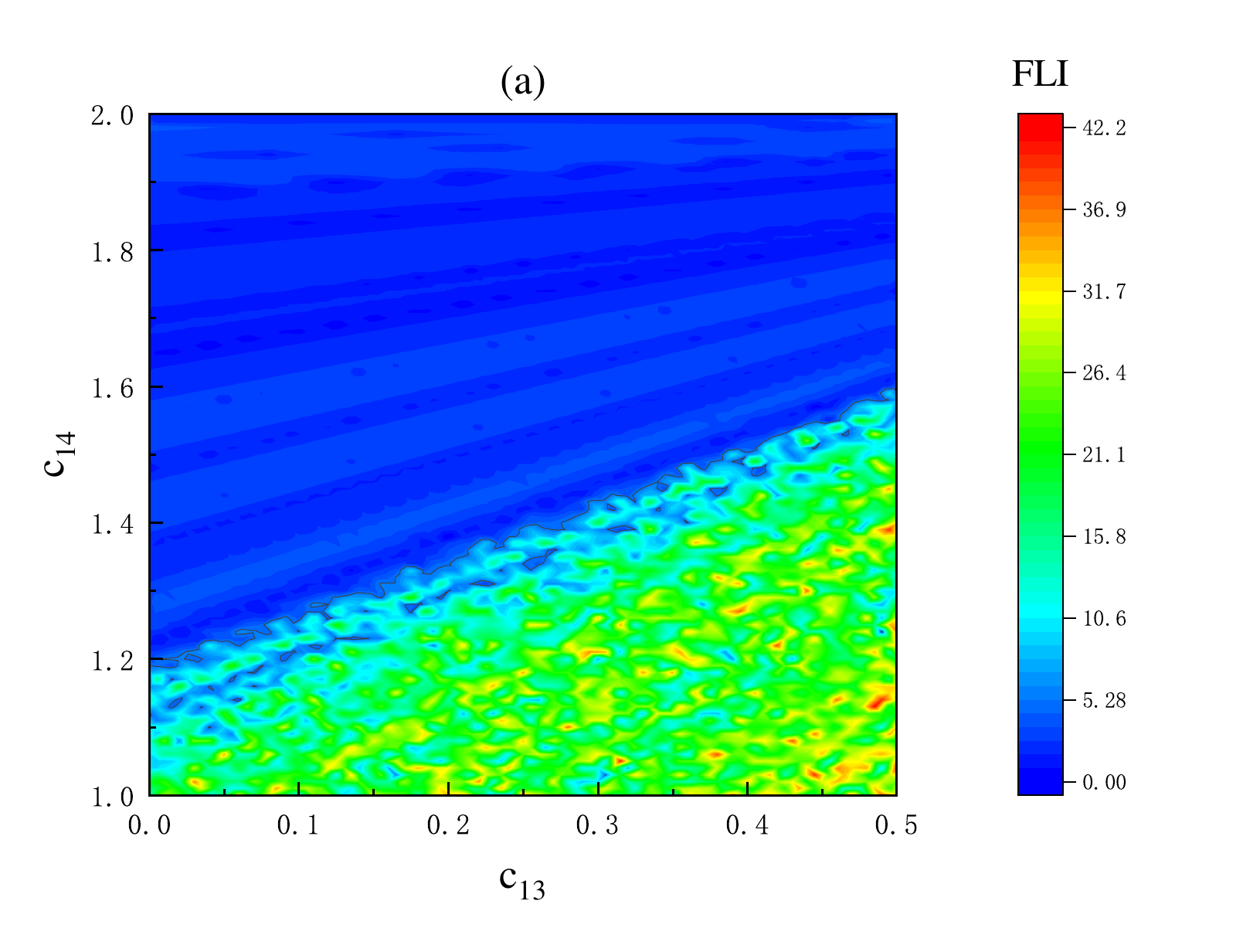}
        \includegraphics[width=15pc]{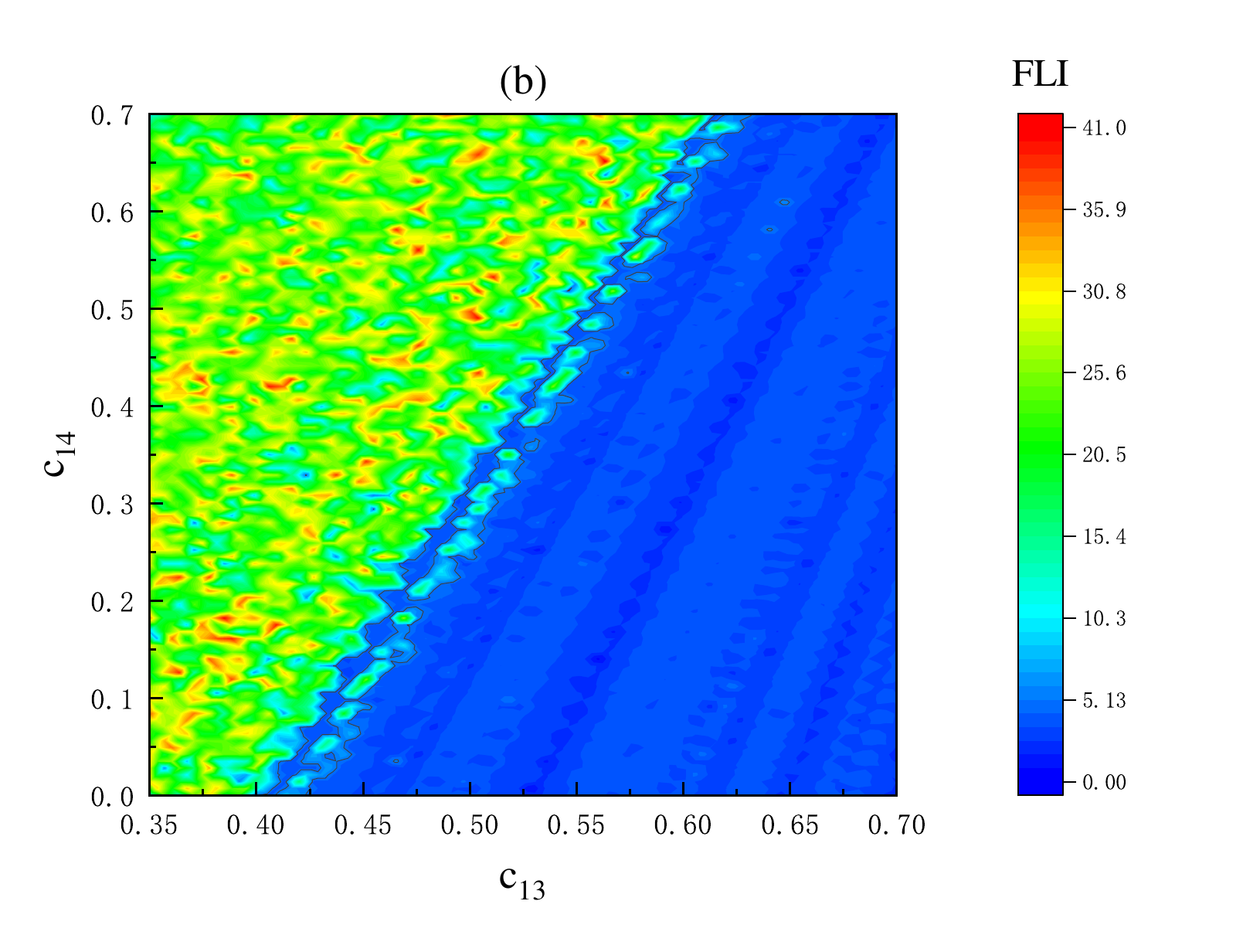}
\caption{Distributions of order and chaos in the two parameter
space $(c_{13}, c_{14})$ using the FLIs.
 The initial radius is $r=35$. (a) $b=9.7\times 10^{-4}$, $E=0.994$ and $L=4.6$ in Case (ii).
(b)$B=9.4\times 10^{-4}$, $E=0.994$ and $L=4.8$ in Case (iii).
        }
    }
\end{figure*}

\begin{figure*}[htpb]
        \centering{
        \includegraphics[width=10pc]{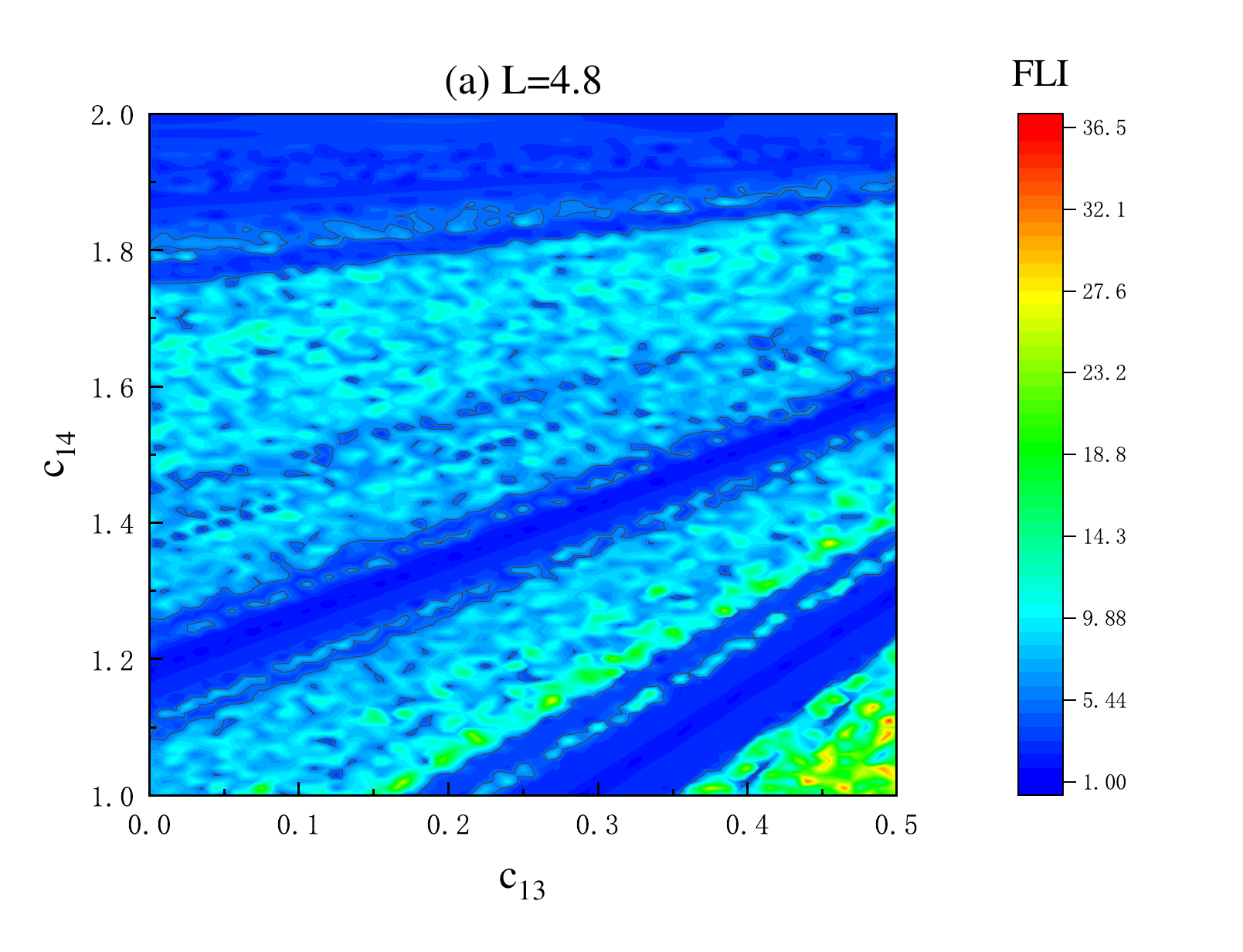}
        \includegraphics[width=10pc]{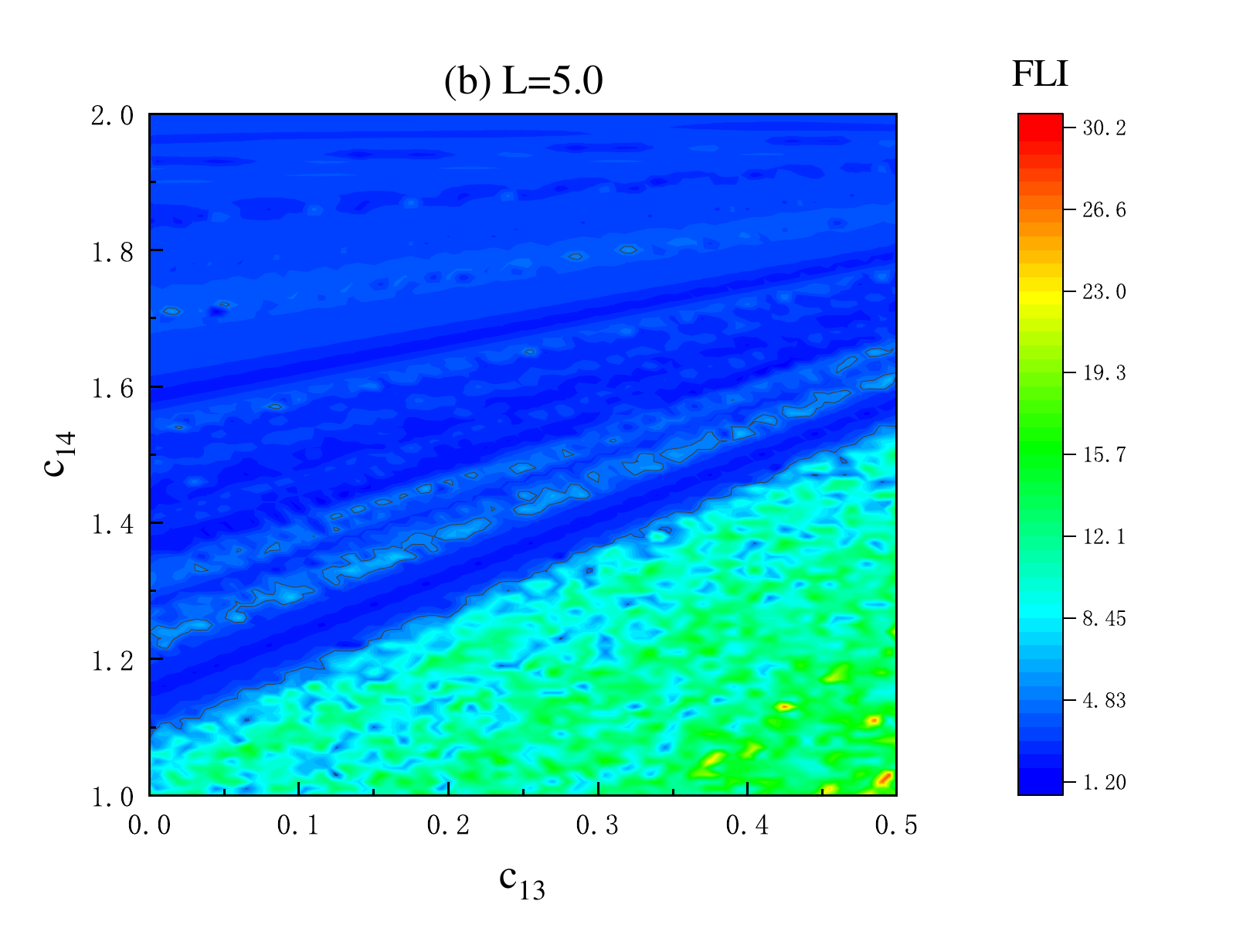}
        \includegraphics[width=10pc]{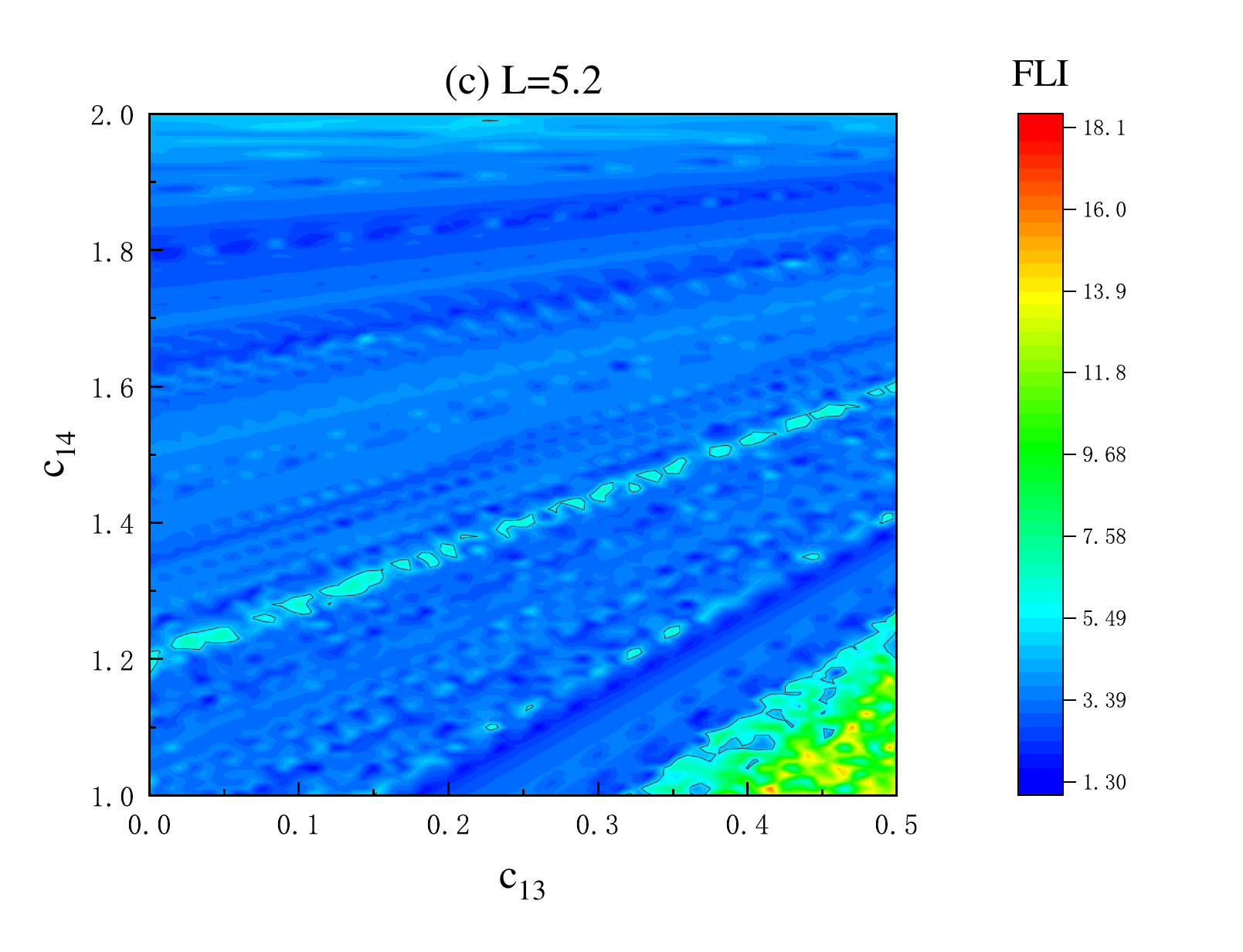}
        \includegraphics[width=10pc]{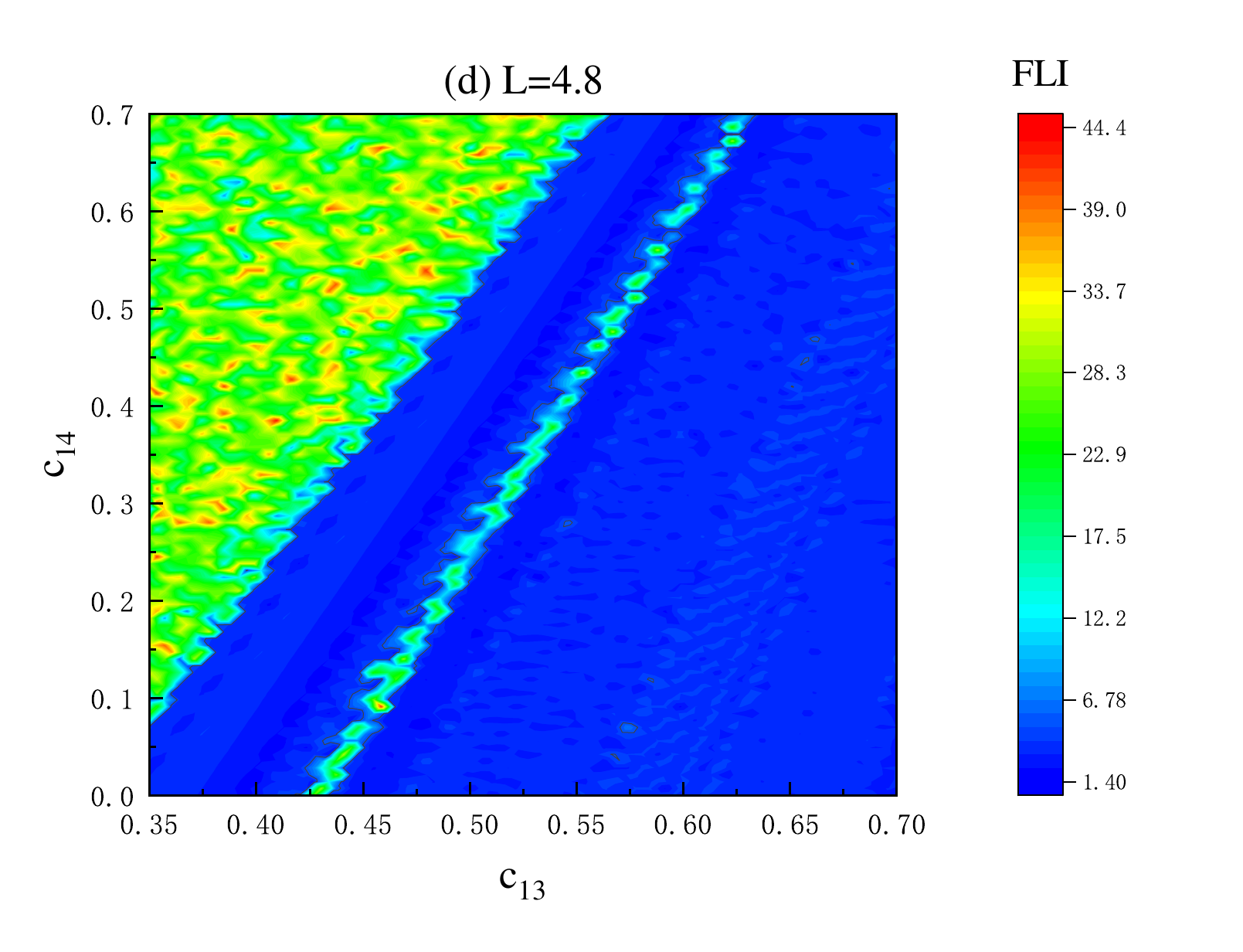}
        \includegraphics[width=10pc]{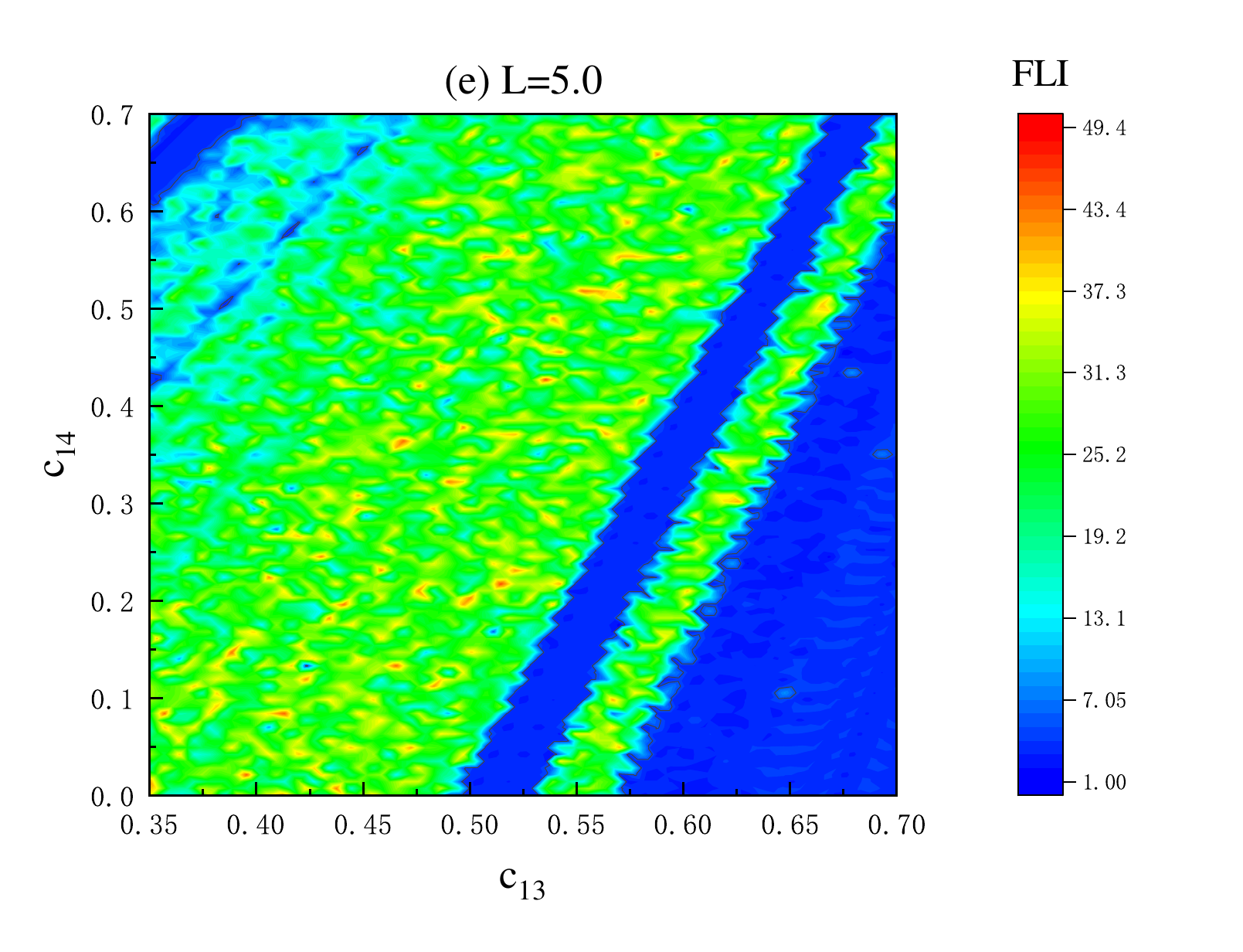}
        \includegraphics[width=10pc]{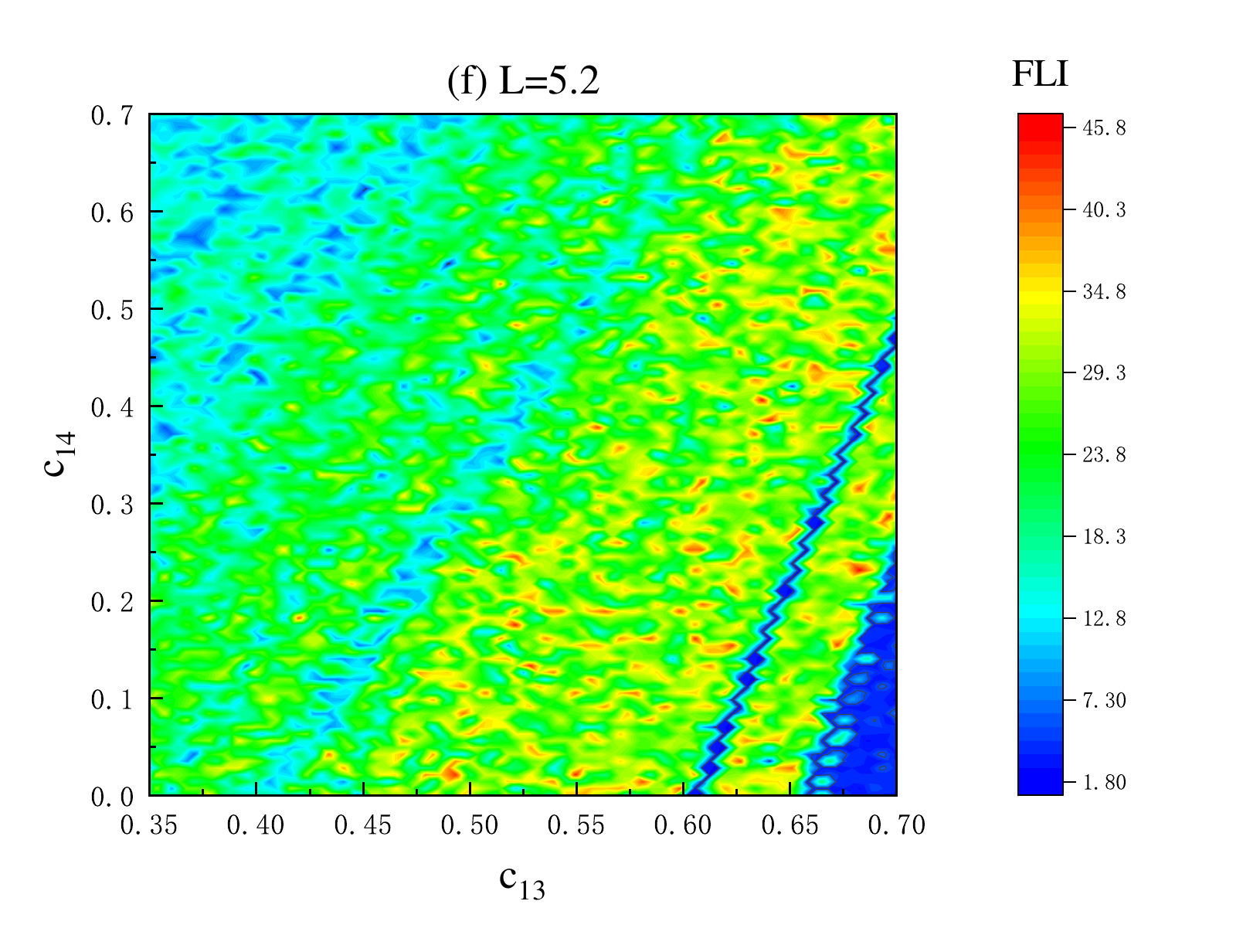}
\caption{Same as Figure 6, but several values are given to the
angular momentum $L$. The initial radius is $r=35$, and the other
parameters are $b=9.1\times 10^{-4}$ and $E=0.995$. (a)-(c): Case
(ii) is considered. (d)-(f): Case (iii) is considered.
                    }
    }
\end{figure*}

\begin{figure*}[htpb]
    \centering{
        \includegraphics[width=10pc]{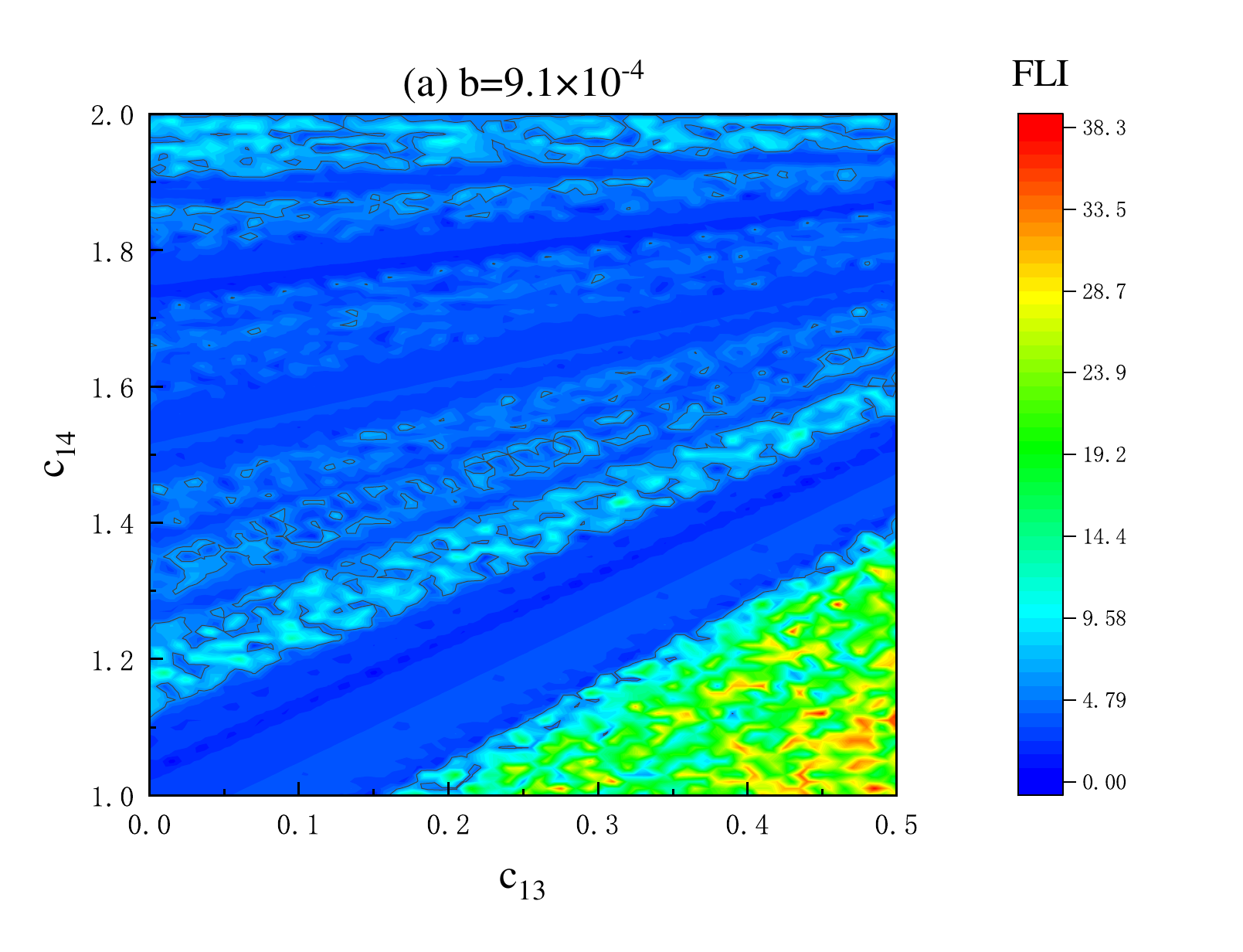}
        \includegraphics[width=10pc]{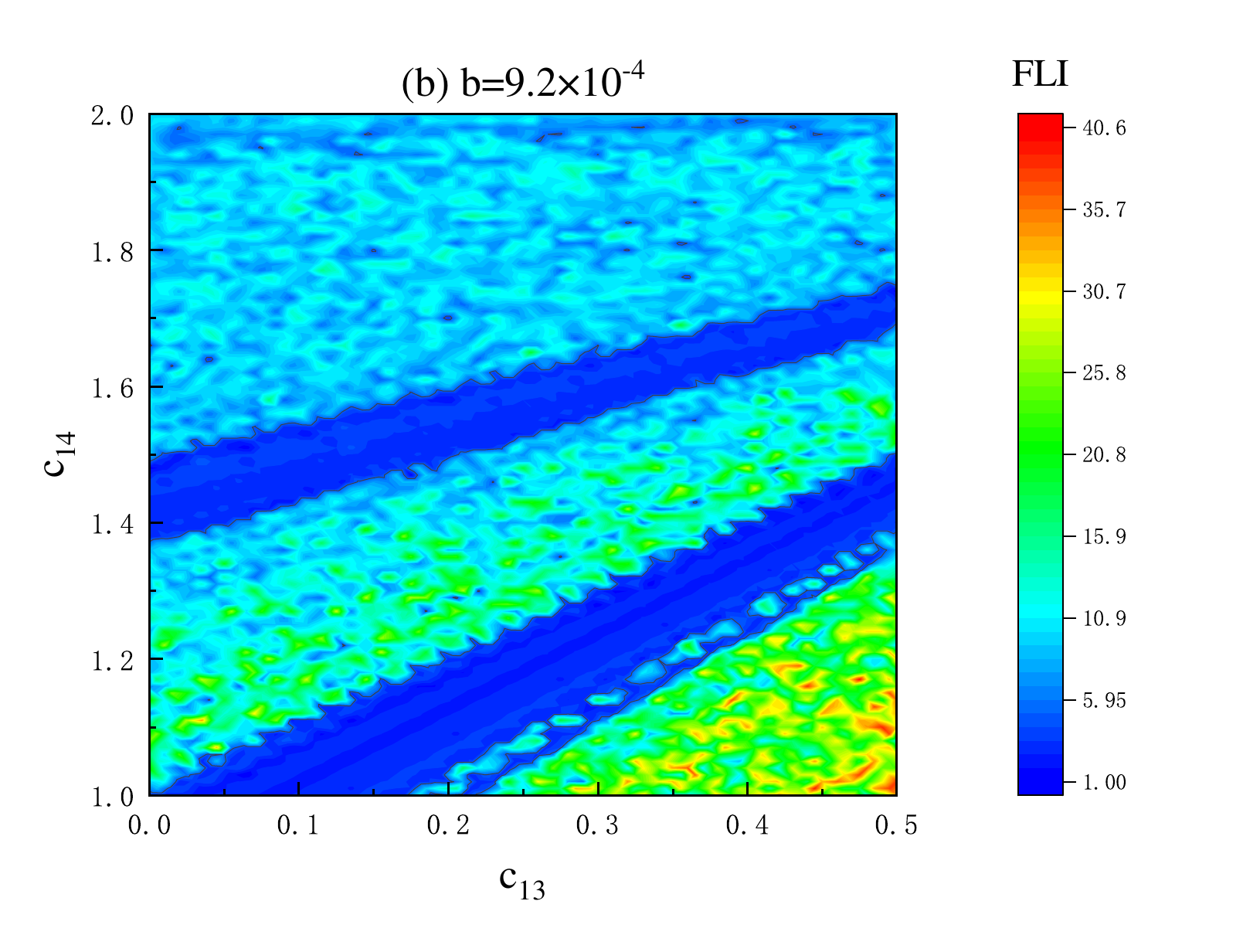}
        \includegraphics[width=10pc]{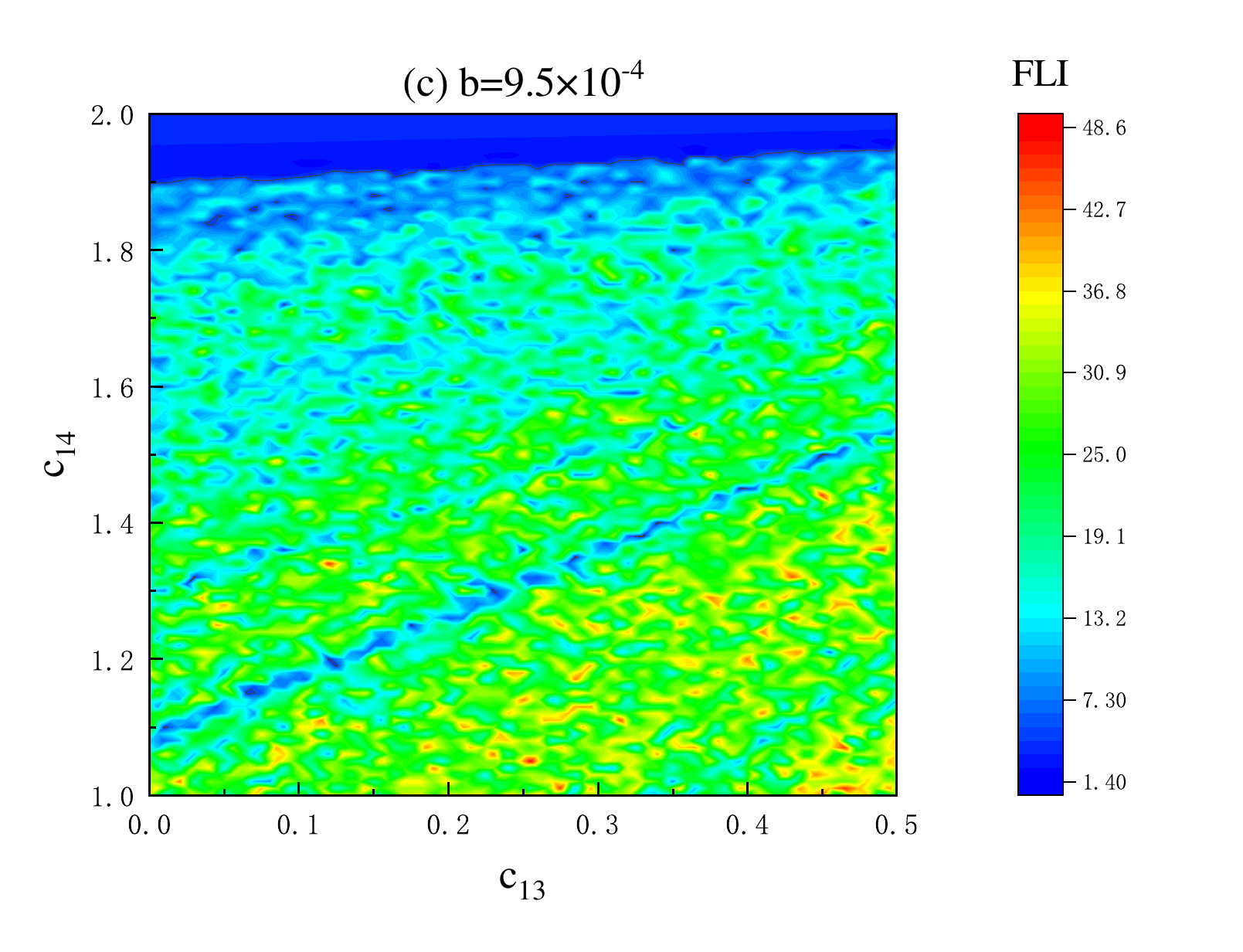}
        \includegraphics[width=10pc]{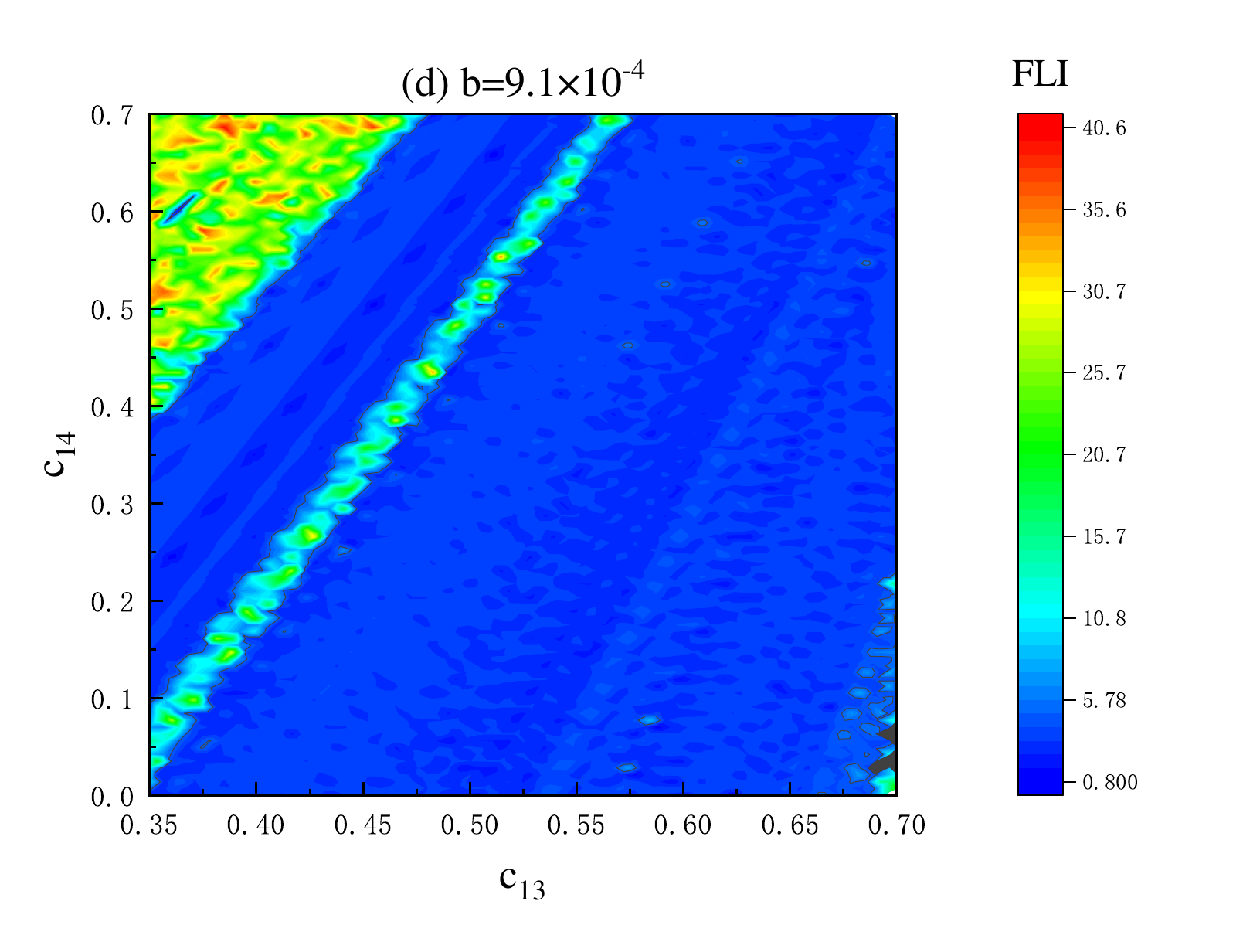}
        \includegraphics[width=10pc]{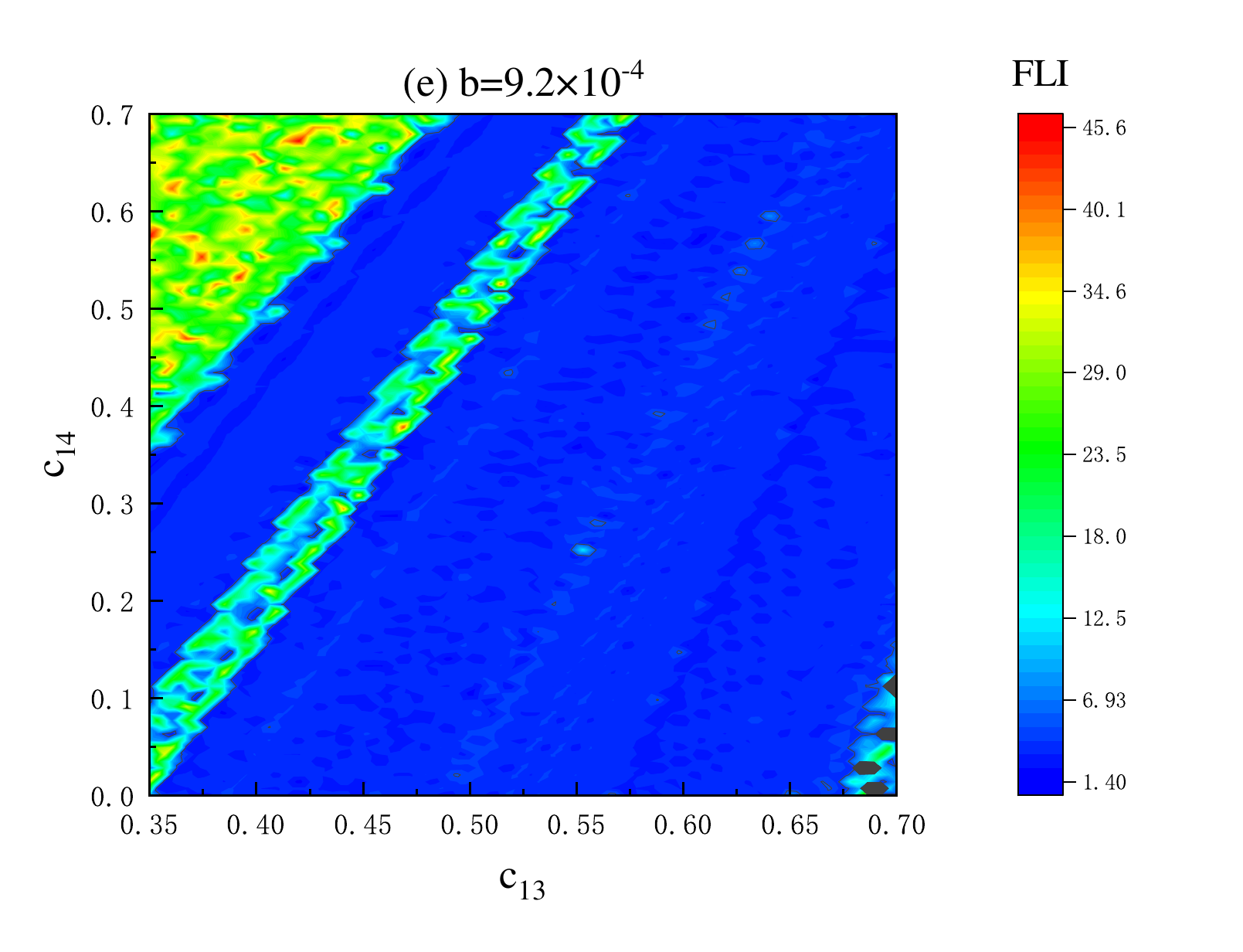}
        \includegraphics[width=10pc]{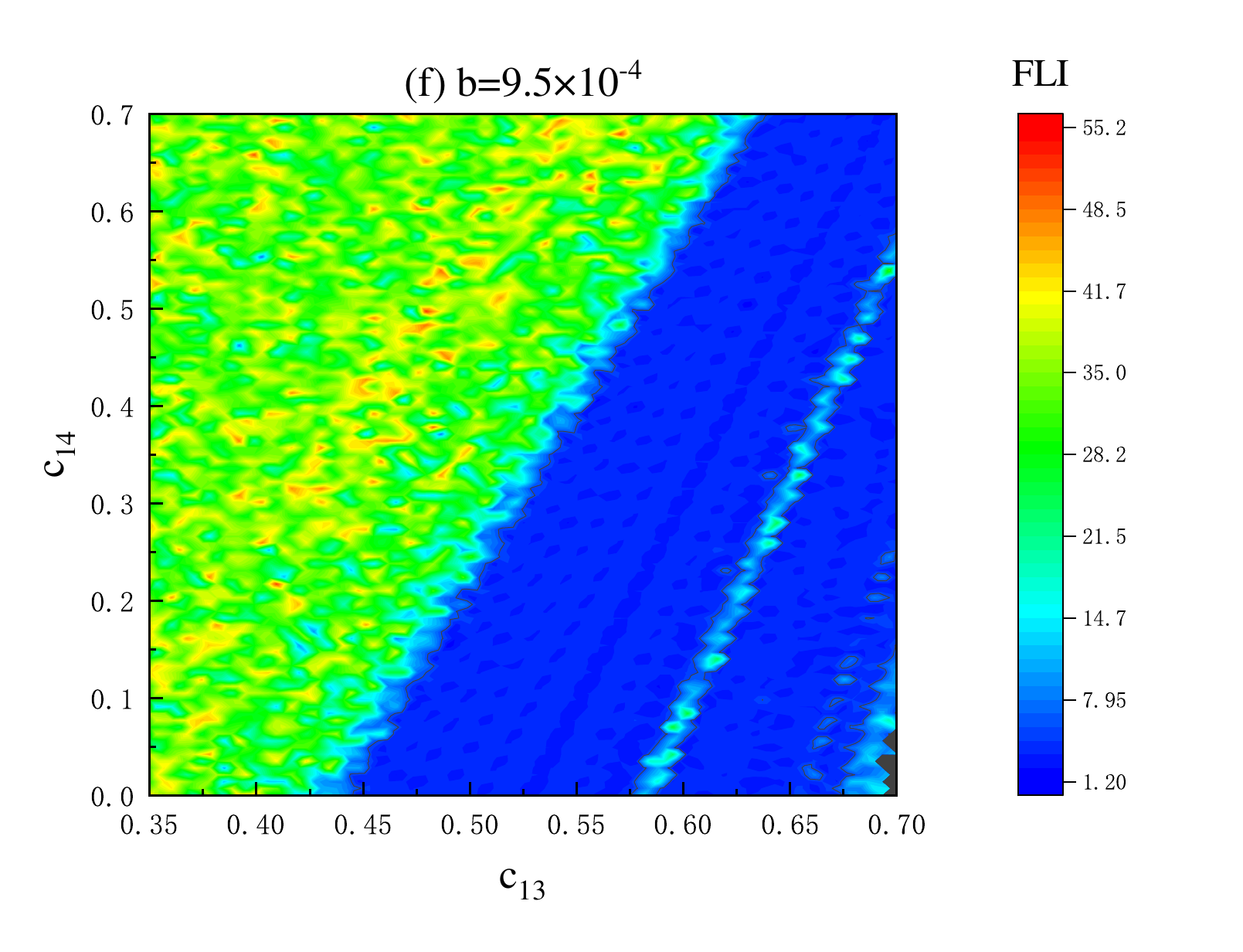}
\caption{Same as Figure 7, but several values are given to the
magnetic field parameter $b$. The initial radius is $r=35$, and
the other parameters are $L=4.7$ and $E=0.995$. (a)-(c): Case
(ii). (d)-(f): Case (iii).
        }
    }
\end{figure*}

\begin{figure*}[htpb]
    \centering{
        \includegraphics[width=10pc]{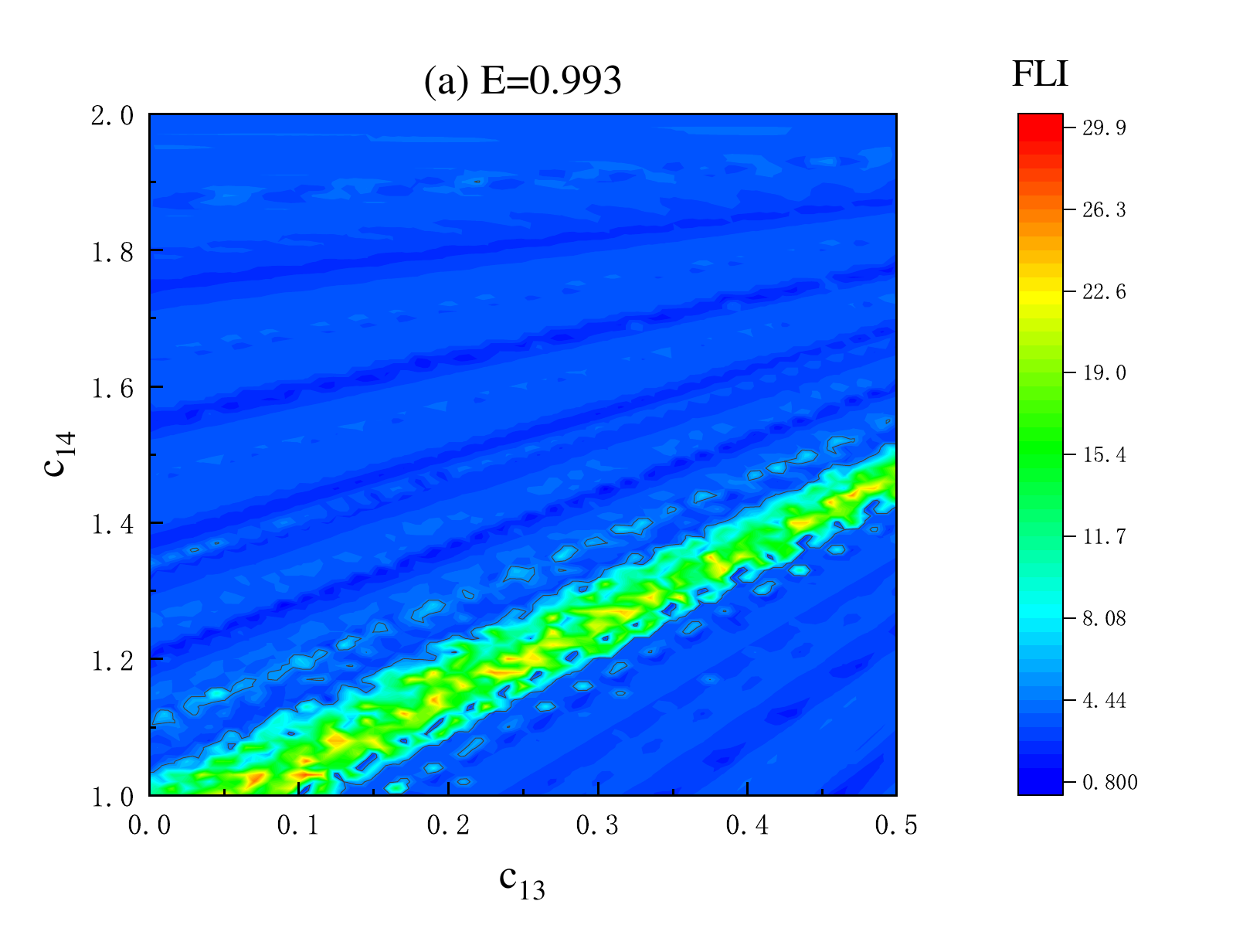}
         \includegraphics[width=10pc]{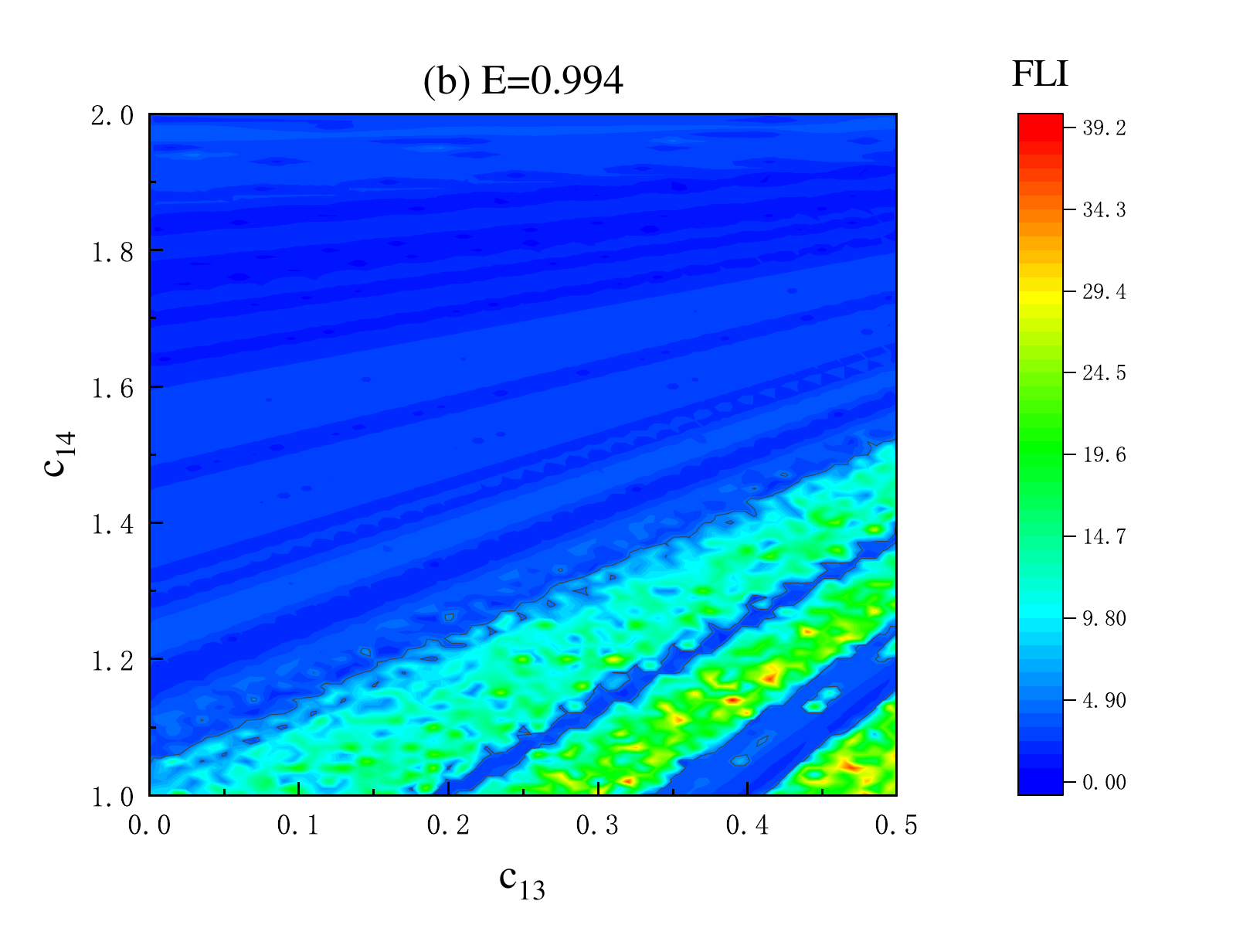}
         \includegraphics[width=10pc]{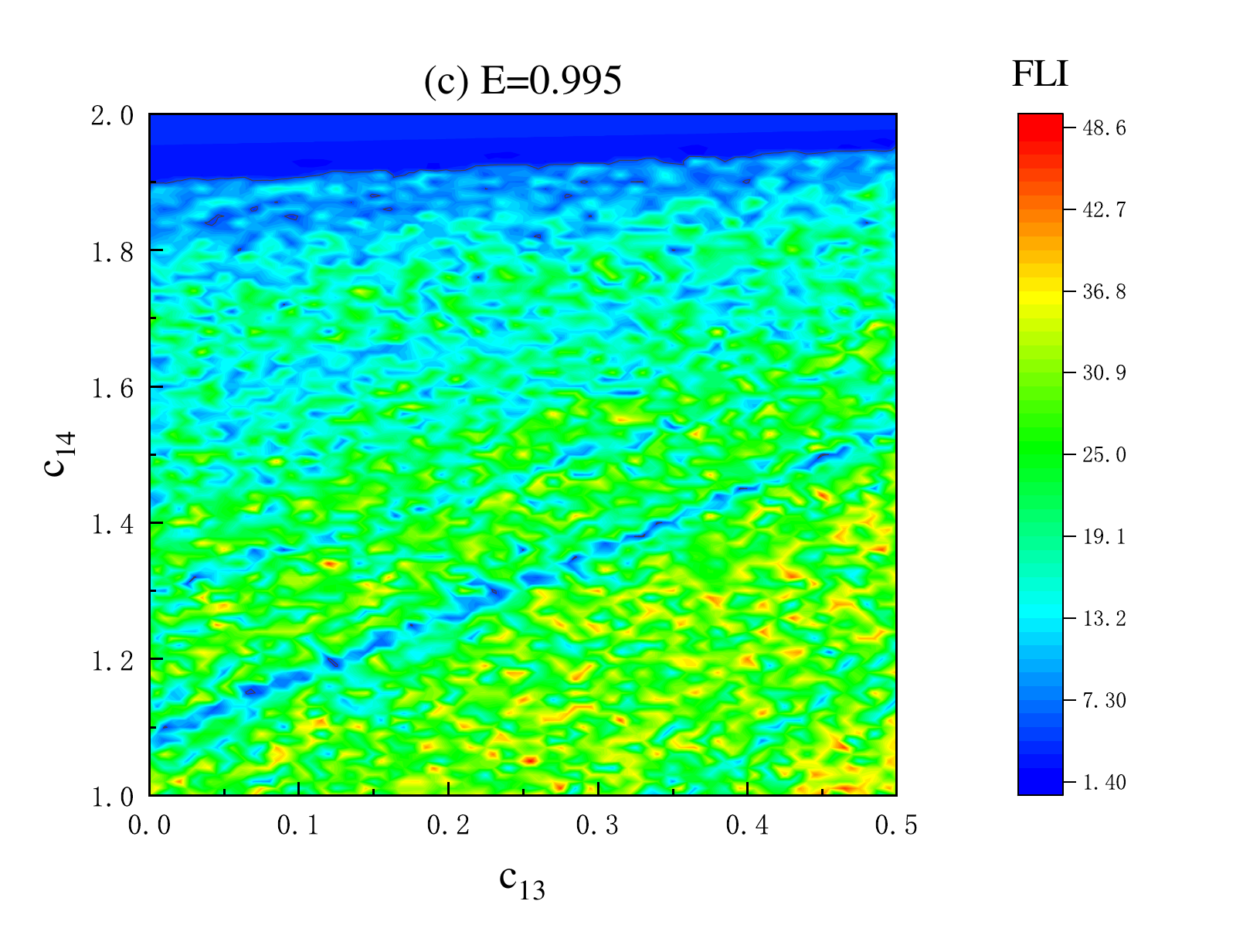}
         \includegraphics[width=10pc]{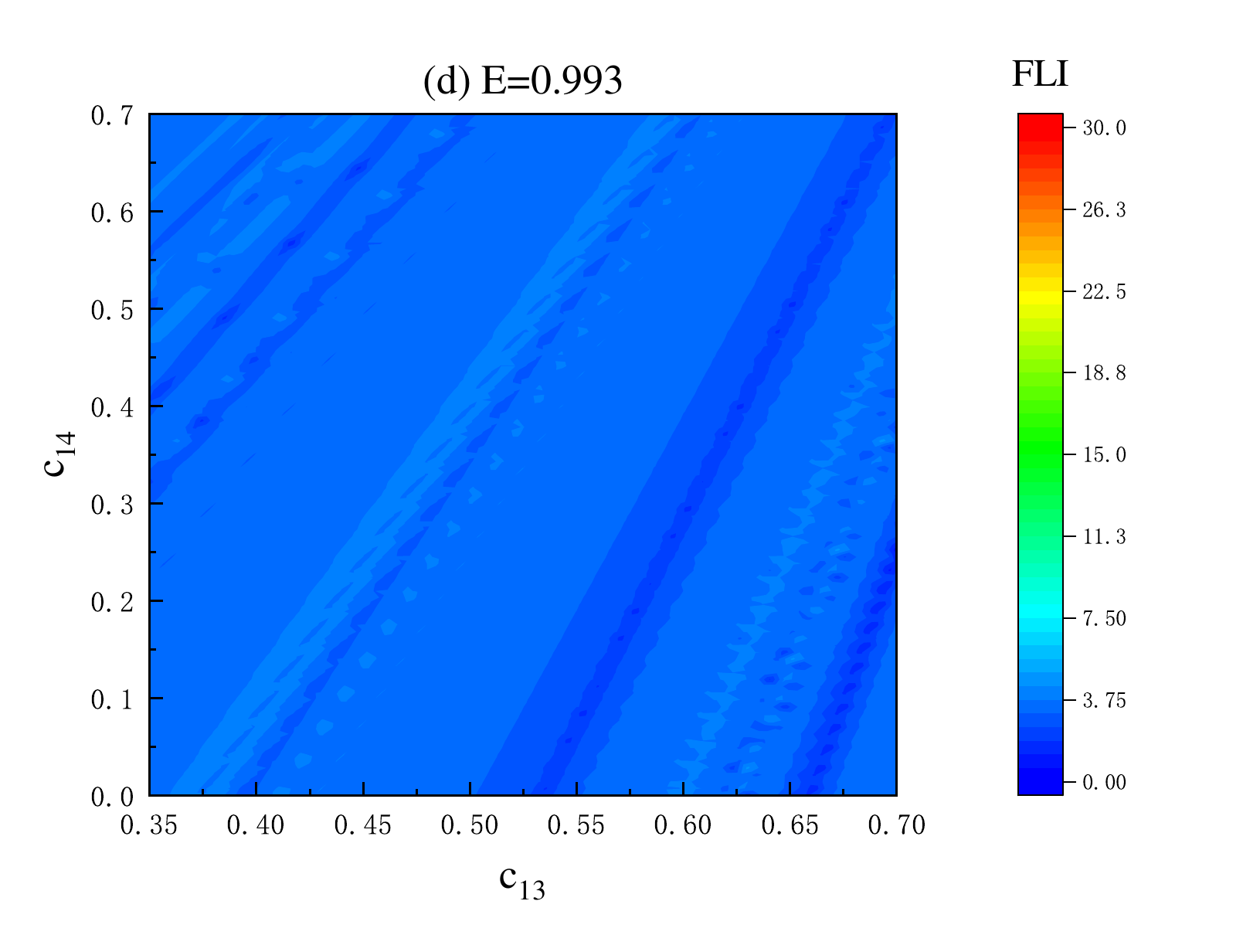}
        \includegraphics[width=10pc]{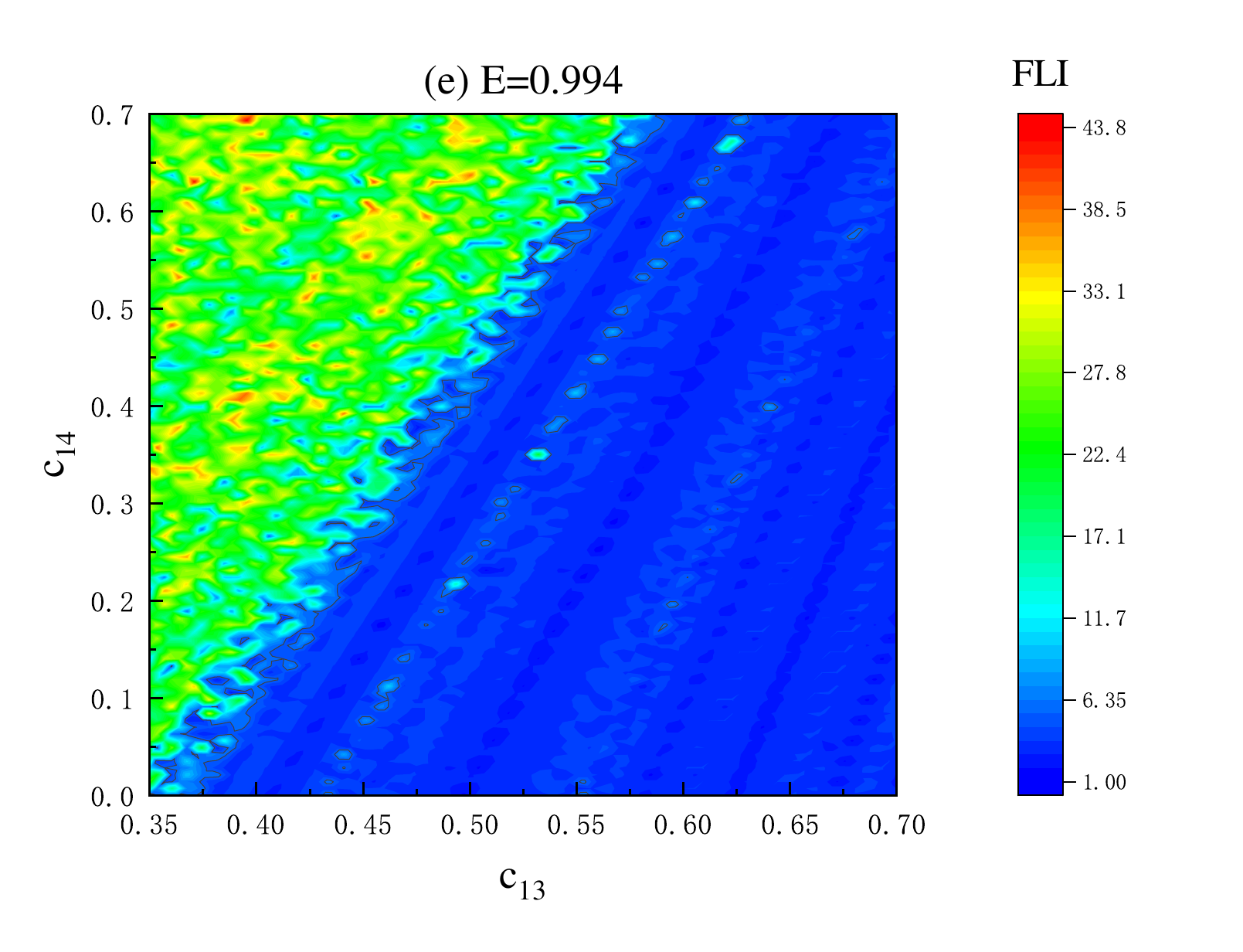}
        \includegraphics[width=10pc]{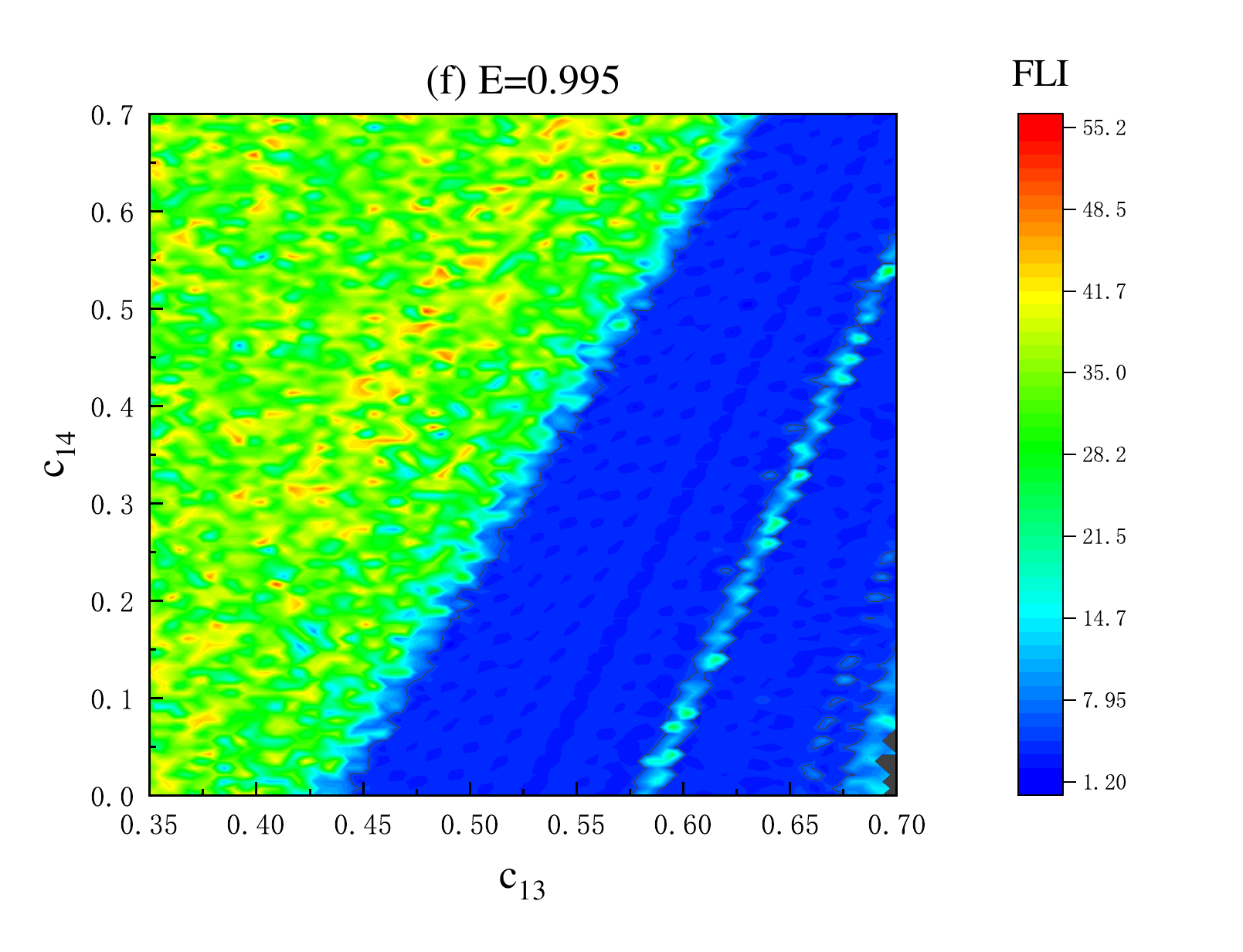}
\caption{Same as Figure 8, but several values are given to the
energy $E$. The initial radius is $r=35$, and the other parameters
are $b=9.5\times 10^{-4}$ and $L=4.7$. (a)-(c): Case (ii).
(d)-(f): Case (iii).
                  }
    }
\end{figure*}

\begin{figure*}[htpb]
    \centering{
        \includegraphics[width=10pc]{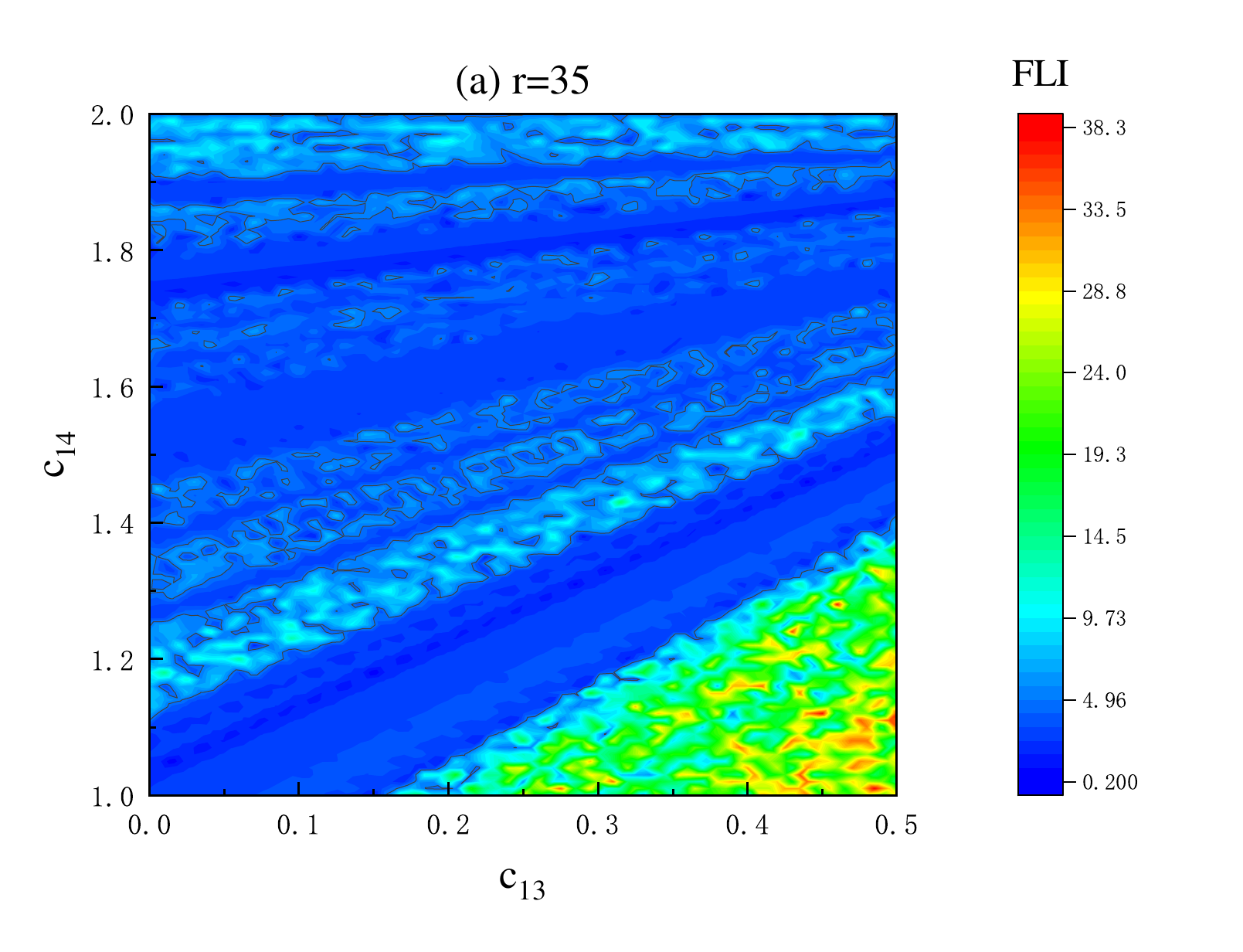}
        \includegraphics[width=10pc]{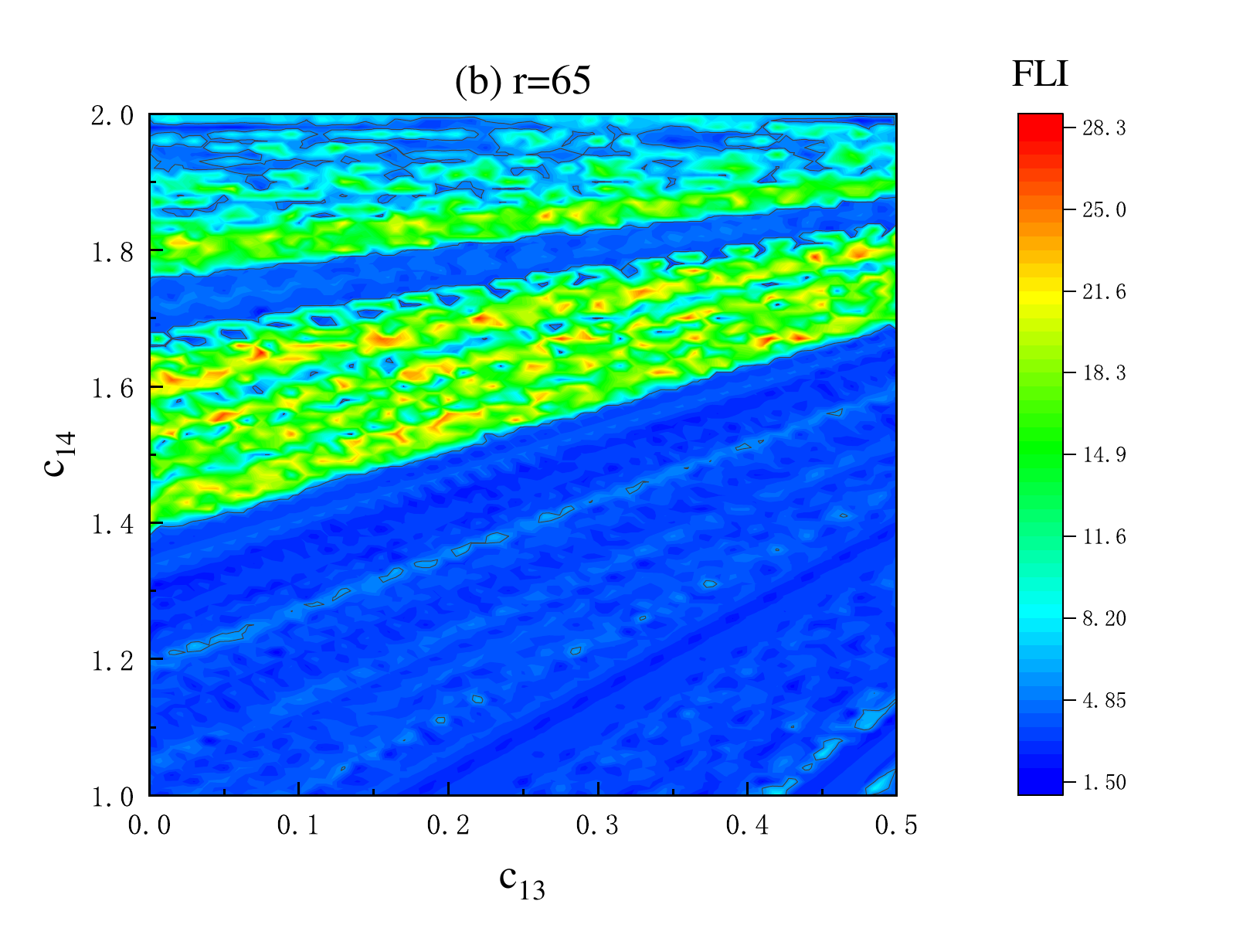}
        \includegraphics[width=10pc]{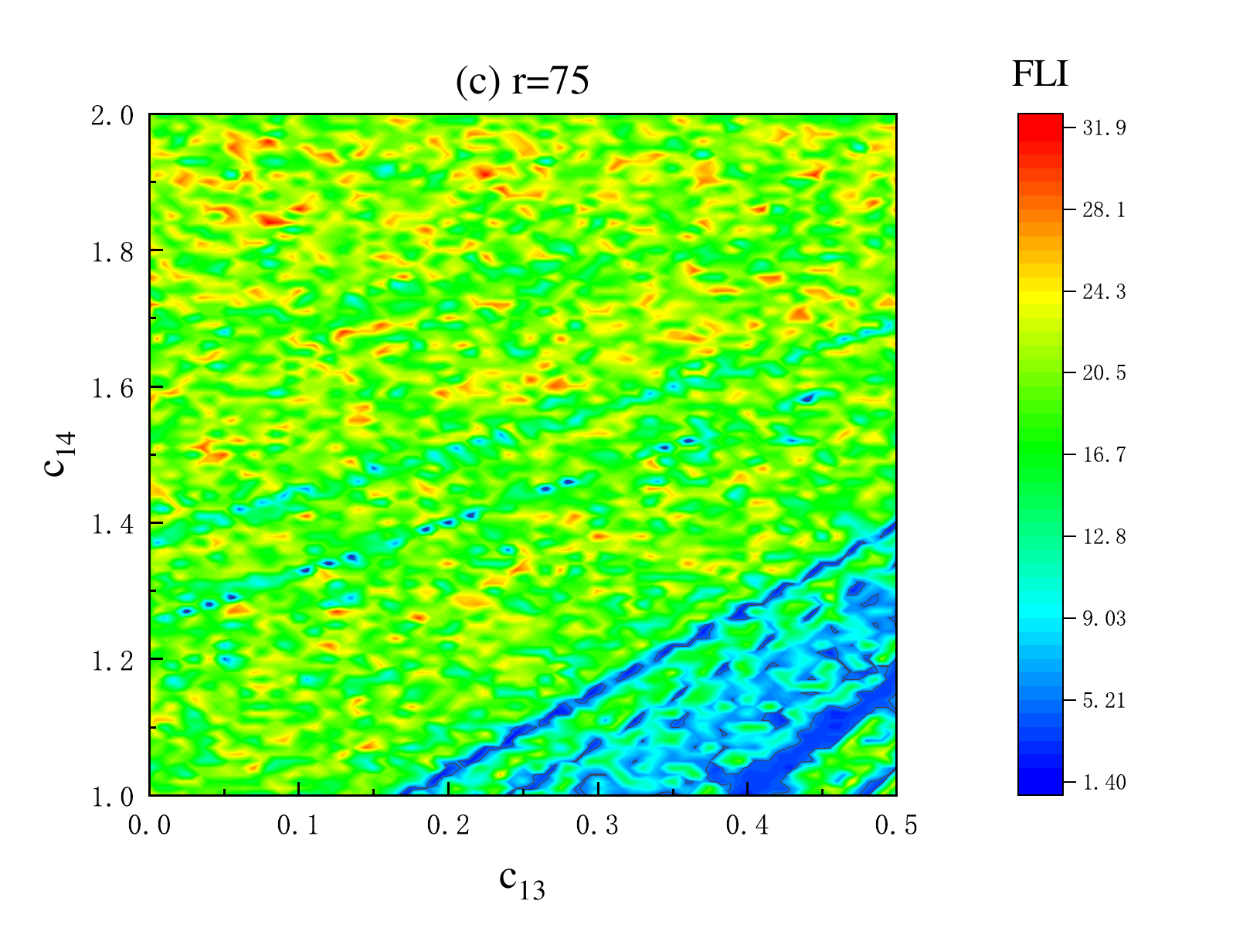}
        \includegraphics[width=10pc]{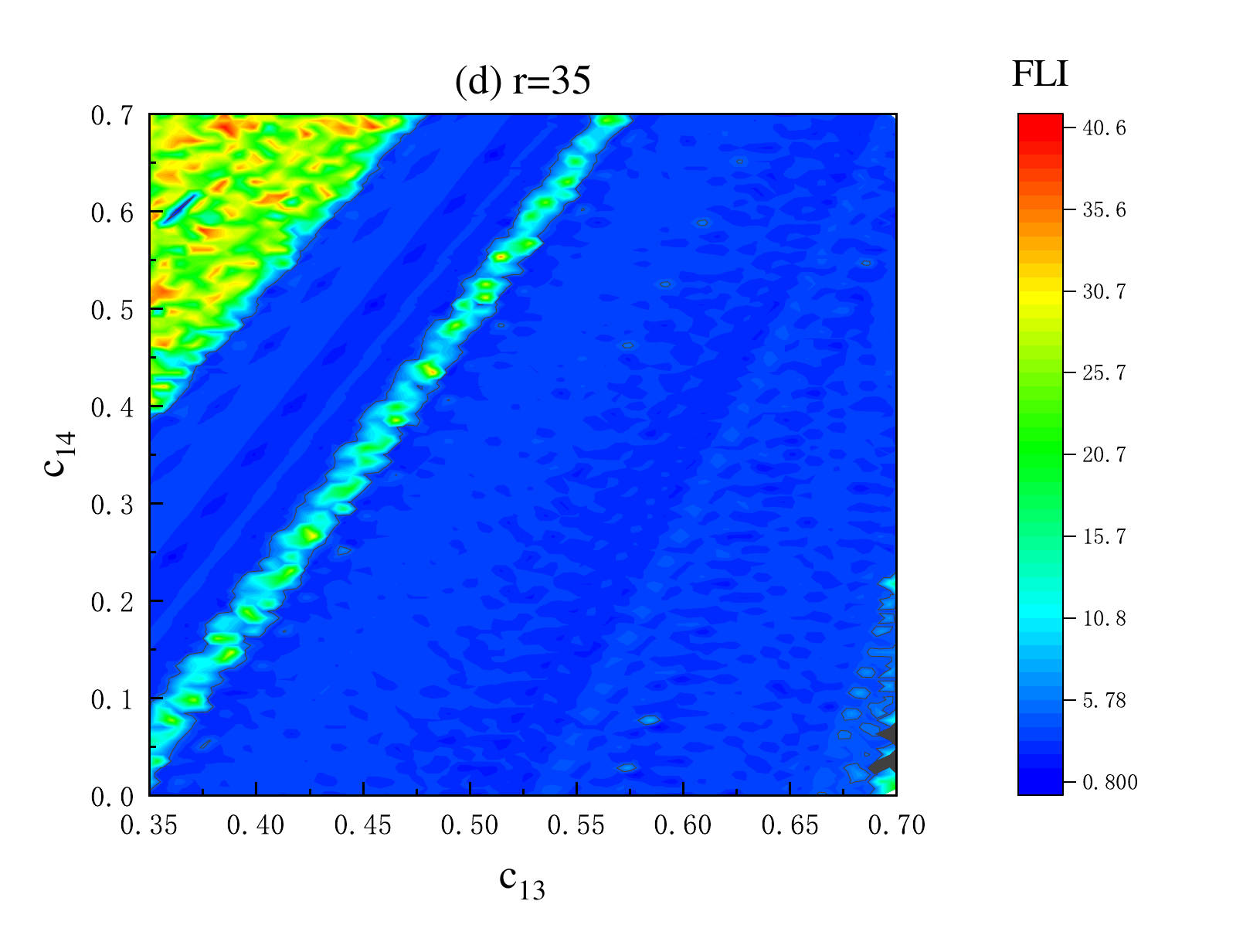}
        \includegraphics[width=10pc]{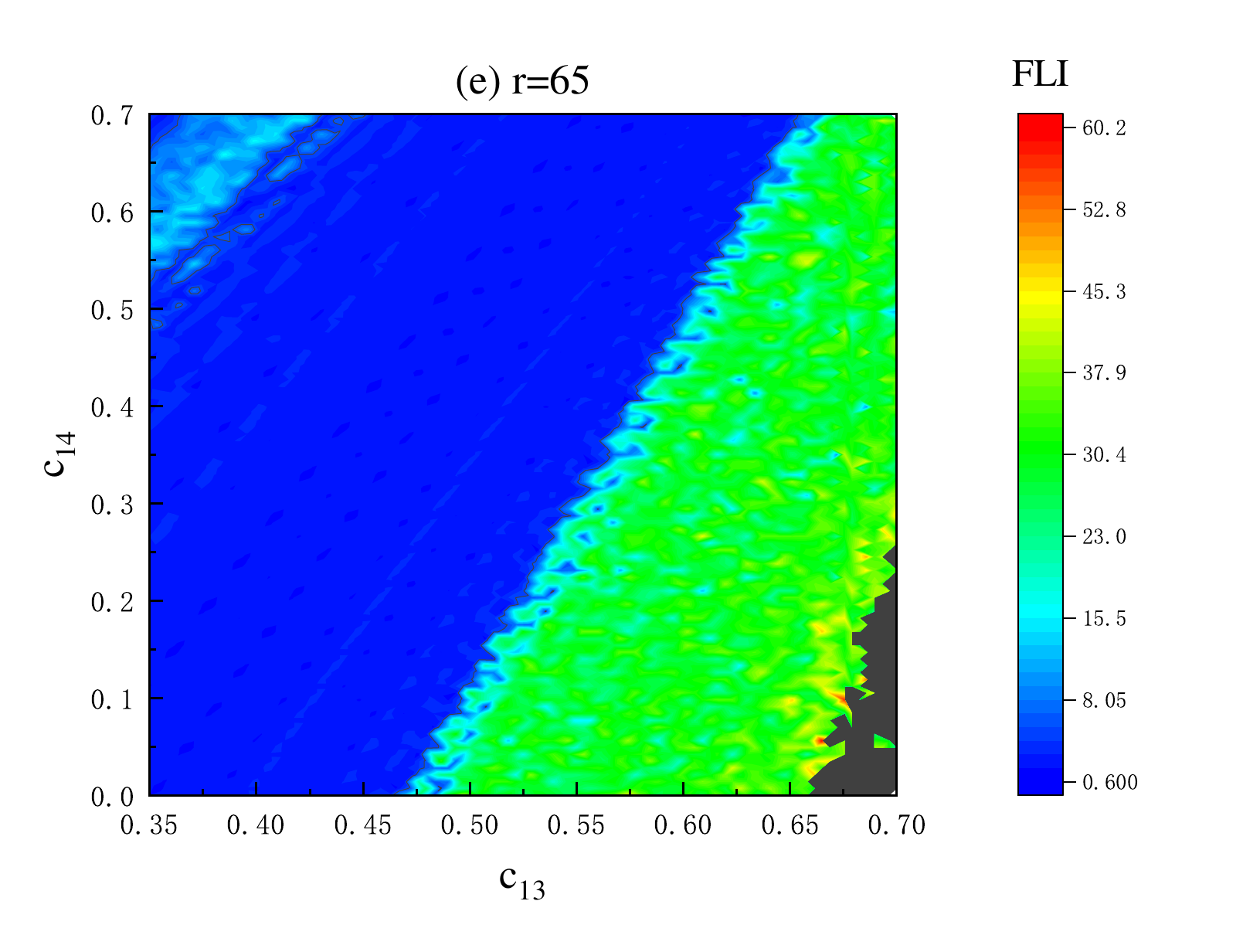}
        \includegraphics[width=10pc]{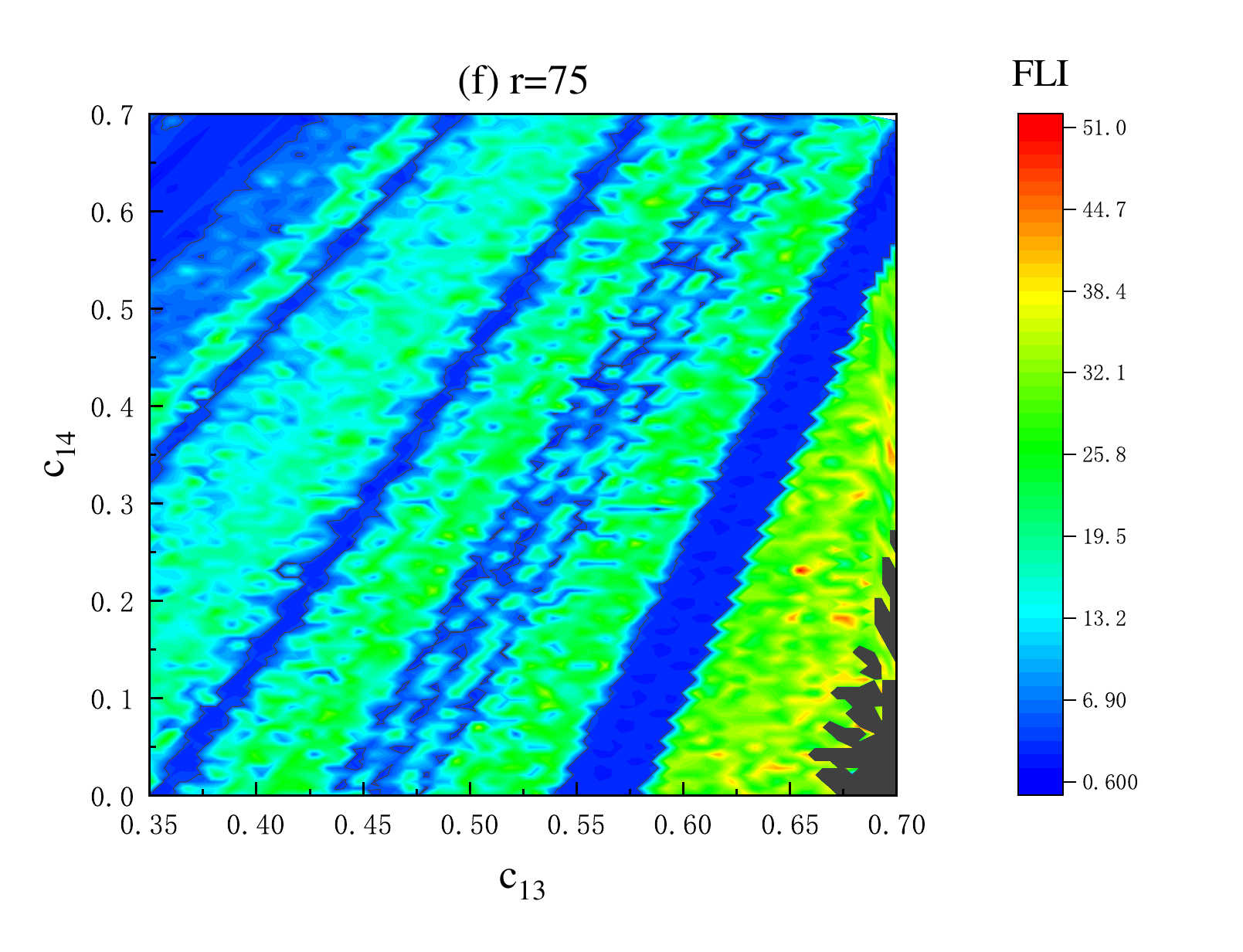}
\caption{Same as Figure 9, but several values are given to the
initial radius $r$. The other parameters are $b=9.1\times
10^{-4}$, $E=0.995$ and $L=4.7$. (a)-(c): Case (ii). (d)-(f): Case
(iii).
                  }
    }
\end{figure*}
\end{document}